\documentclass[aps,prb,10pt,twocolumn,amsmath,amssymb,longbibliography,superscriptaddress,nofootinbib]{revtex4-2}
\usepackage{amsmath,amssymb,amsfonts} 	
\usepackage{graphicx}
\usepackage{hhline}
\usepackage[latin1]{inputenc}
\usepackage{subfigure}
\usepackage[normalem]{ulem}
\usepackage{xcolor}
\usepackage[version=4]{mhchem} 
\begin{document}

\title{Abundance of Second Order Topology in C$_3$ Symmetric Two-dimensional Insulators}

\author{Joachim S\o dequist}
\affiliation{CAMD, Department of Physics, Technical University of Denmark, 2820 Kgs. Lyngby Denmark}
\author{Urko Petralanda}
\affiliation{CAMD, Department of Physics, Technical University of Denmark, 2820 Kgs. Lyngby Denmark}
\author{Thomas Olsen}
\email{tolsen@fysik.dtu.dk}
\affiliation{CAMD, Department of Physics, Technical University of Denmark, 2820 Kgs. Lyngby Denmark}

\begin{abstract}
We have screened 71 two-dimensional (2D) materials with $C_3$ symmetry for non-trivial second order topological order and find that 28 compounds exhibit an obstructed atomic limit (OAL). In the case of $C_3$ symmetry, the second order topology can be calculated from bulk symmetry indicator invariants, which predict the value of fractional corner charges in symmetry conserving nanoflakes. The procedure is exemplified by MoS$_2$ in the H-phase, which constitutes a generic example of a 2D OAL material and the predicted fractional corner charges is verified by direct calculations of nanoflakes with armchair edges. We also determine the bulk topological polarization, which always lead to gapless states at zigzag edges and thus deteriorates the concept of fractional corner charges in nanoflakes with zigzag edges that are typically more stable that armchair flakes. We then consider the case of TiCl$_2$, which has vanishing polarization as well as an OAL and we verify that the edge states of nanoflakes with zigzag edges may indeed by passivated such that the edges remain insulating and the corner charges are well defined. For the 28 OAL materials we find that 16 have vanishing polarization and these materials thus constitute a promising starting point for experimental verification of second order topology in a 2D material.
\end{abstract}
\maketitle

\section{Introduction}
The discovery of the quantum spin Hall effect in 2005 \cite{Kane2005,Kane2005a} has initiated an immense interest in the topological properties of crystalline solids \cite{Bernevig2006, Konig2007, Hasan2010, Moore2010, Brune2012, Qian2014, Olsen2018}. The initial $\mathbb{Z}_2$ classification of time-reversal invariant two-dimensional (2D) insulators was rapidly generalized to three-dimensional (3D) insulators \cite{Fu2007, Fu2007a,  Qi2008, Zhang2009} and in 2011 it was shown that non-trivial topology may be protected by crystalline symmetries as well \cite{Fu2011, Hsieh2012, Kargarian2013}. A common signature of such "first order" topological insulators is the appearance of gapless states that are guaranteed to exist at symmetry-conserving edges and surfaces in 2D and 3D materials respectively \cite{Brune2012, Liu2014, Fei2017, Ugeda2018}. 

In addition to the topological classification scheme based on time-reversal symmetry and crystalline symmetries, certain non-polar space groups allow for quantized values of the spontaneous polarization, which can thus be regarded as a topological index in itself \cite{Benalcazar2017a}. 

The simplest example is comprised by the one-dimensional Su-Schrieffer-Heeger (SSH) chain \cite{Su1979}, which yields a polarization of either 0 or $e/2$. The two phases correspond to Wannier charge centers being localized at inversion symmetric lattice sites or on bonds between sites. The latter case is referred to as a obstructed atomic limit (OAL) because the system cannot be adiabatically dissociated without breaking the symmetry.
In a chain with open boundary conditions that preserve the symmetry and charge neutrality, the OAL phase will have Wannier charge centers at the end points that cannot both be occupied implying that the symmetry is either spontaneously broken or that the degenerate edge states are partially occupied. Alternatively, one may relax the condition of charge neutrality and the edge states will be fully occupied with a resulting fractional boundary charge of $e/2$. The concept of topological polarization is readily generalized to higher dimensions and the ubiquitous metallic edge states in nanoparticles of 2D MoS$_2$ have been attributed to a non-trivial topological polarization \cite{Gibertini2015}. However, in 2D it is possible to obtain a different OAL phase where the edges are insulating, but carry a finite dipole moment that results in fractional corner states \cite{Benalcazar2017a,Benalcazar2017b, Benalcazar2019, Song2017b, Langbehn2017b, Schindler2018a}, and the topology is then referred to as being of second order.
In 3D, non-trivial second order topology gives rise to gapless hinge states that has been observed in bulk Bi \cite{Schindler2018b}, but so far there is no direct experimental evidence for second order topology and the resulting fractional corner charges in a 2D material.
\begin{figure}[tb]
        \centering
        \includegraphics[width=0.4\textwidth]{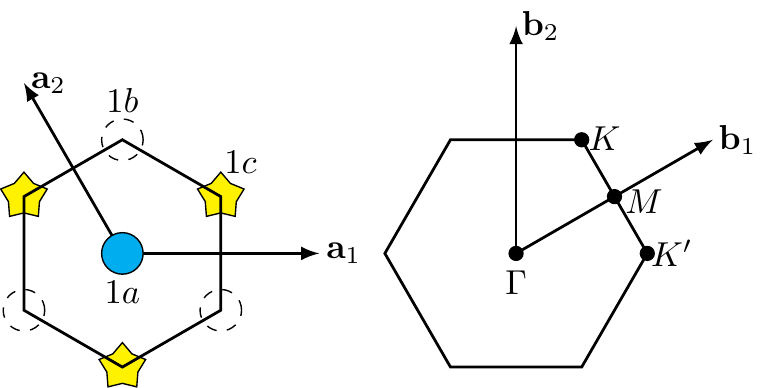}
        \caption{Left: unit cell of MoS$_2$ with maximal Wyckoff positions. Mo is indicated by blue circle and S is indicated by yellow stars. Right: Brillouin zone of MoS$_2$ with indication of the high symmetry points $\Gamma$, $M$, $K$ and $K'$.}
        \label{fig:unitcell}
\end{figure}

In the present letter we have screened 71 2D materials with $C_3$ symmetry for non-trivial second order topology. Of these we find that 28 is in the OAL phase, which thus appears to be a rather common feature of 2D materials. We start by analyzing 2D MoS$_2$ in the H-phase, which comprises a prototypical example of a material which can be characterized by an OAL phase as well as by a topological polarization in the armchair direction \cite{Gibertini2015, Qian2021, Zeng2021}. The OAL phase can be predicted from the bulk material and the emergence of fractional corner charges in symmetry-conserving flakes with armchair edges is explicitly verified. It is also shown that the topological polarization of MoS$_2$ yields metallic edges at any zigzag edge in agreement with previous predictions. We then consider the case of TiCl$_2$, which also exhibits an OAL phase, but has vanishing polarization and we show that in that case all nanotriangles exhibit fractional corner charges and edges that are either insulating or may be made so by including appropriate edge adsorbates. Thus we argue that in absence of metallic edge states, the localized corner states should be experimentally accessible using atomically resolved scanning tunneling microscopy (STM) \cite{helveg2000atomic}. Finally, we summarize our results on the 28 OAL materials and we find that 16 of these have vanishing polarization, which implies candidacy for experimental verification of second order topology in 2D.

\section{Topological properties of H-MoS$_2$}
\begin{figure}[tb]
        \centering
        \includegraphics[width=0.45\textwidth]{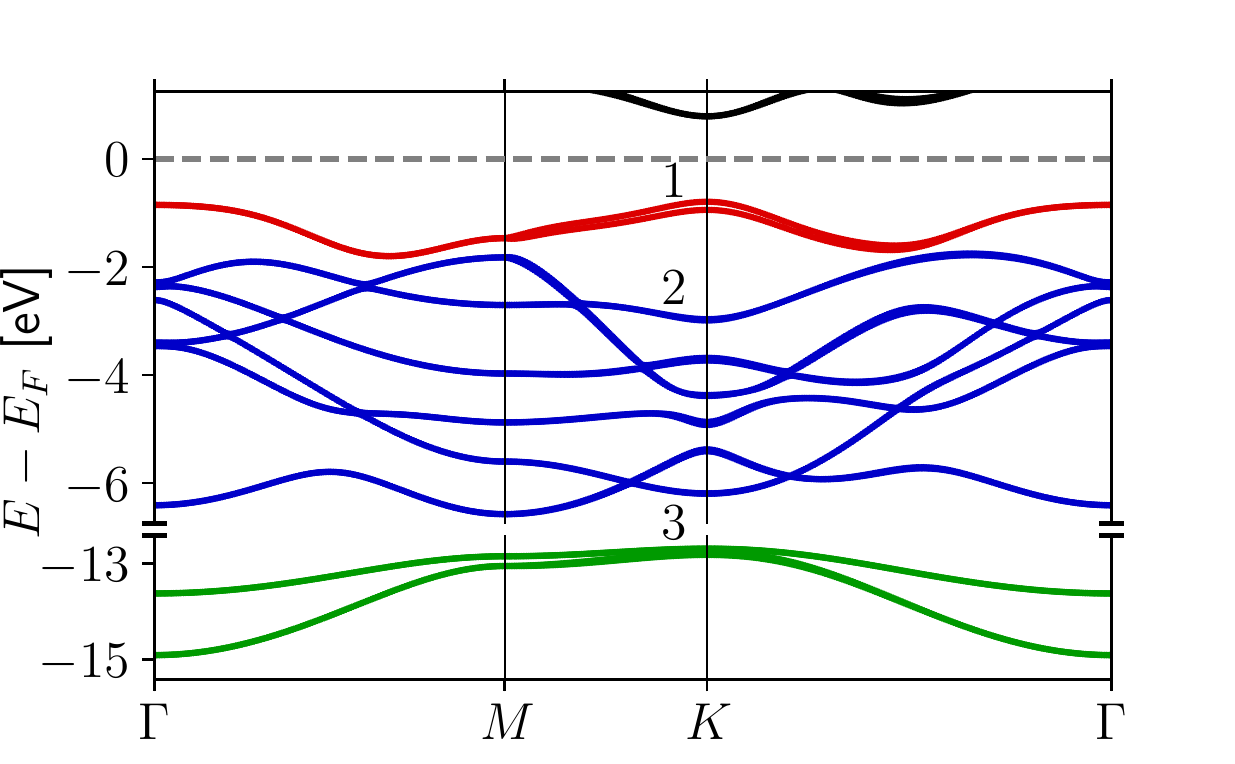}
        \caption{Band structure of MoS$_2$. The three groups of bands correspond to occupancy of different Wyckoff positions as summarized in Tab. \ref{tab:MoS2invariants}.}
        \label{fig:bandstructure}
\end{figure}
The unit cell of 2D MoS$_2$ is shown in Fig. \ref{fig:unitcell} where the atomic positions and maximal Wyckoff positions are indicated. The OAL phase is related to the occupancy of a Wannier function at the $1b$ site, which may give rise to non-trivial fractional corner charges. The occupation of the maximal Wyckoff positions can be calculated from the symmetry indicator invariants \cite{Schindler2019}
\begin{equation}
    \chi^{(3)} = \{[K^{(3)}_1],[K^{(3)}_2]\},
    \label{eq:symmetry_indicator}
\end{equation}
where $[K^{(3)}_i]$ is the difference in the number of occupied $C_3$ rotation eigenvalues between $K$ and $\Gamma$ (see Fig. \ref{fig:unitcell}). Here $i=1$ refers to the rotation eigenvalue $e^{i\pi/3}$ and $i=2$ refers to the rotation eigenvalue $-1$. The symmetry indicator \eqref{eq:symmetry_indicator} in itself constitutes a topological invariant, but depends on the choice of unit cell and non-trivial values (different from (0, 0)) do not in general describe OAL phases. Instead, the OAL phase can be related to appearance of fractional corner charges which can be calculated from \cite{Schindler2019}
\begin{equation}
    Q_\mathrm{c}(C_3) = \frac{2e}{3}([K_1^{(3)}]+[K_2^{(3)}]) \mod 2e,
    \label{eq:C3corner}
\end{equation}
and thus comprises a $\mathbb{Z}_3$ classification of the second order topology. In addition, if the unit cell is chosen with the origin at a three-fold rotation axis the electronic part of the 2D polarization becomes quantized according to
\begin{equation}
    \textbf{P}_\mathrm{el}^{(3)} = \frac{e}{3A}\Big(2[K_1^{(3)}]+[K_2^{(3)}]\Big)(2\textbf{a}_1+\textbf{a}_2) \mod e\mathbf{R}_j/A,
    \label{eq:C3polarization}
\end{equation}
where $A$ is the unit cell area, $\mathbf{R}_j$ is an arbitrary lattice vector and the lattice constants are chosen as $\textbf{a}_1 = a\hat{\textbf{x}}$ and $\textbf{a}_2 = -\frac{1}{2} a\hat{\textbf{x}}+\frac{\sqrt{3}}{2}a\hat{\textbf{y}}$ with the lattice constant $a$. The polarization thus constitutes a distinct $\mathbb{Z}_3$ topological index that in dimensionless units may take values of $(0, 0)$, $(2/3, 1/3)$ and $(1/3, 2/3)$, which are just the high symmetry points of the Wigner-Seitz cell. For a single band these values thus correspond to Wannier charge centers located at $1a$, $1b$ or $1c$ respectively. 
\begin{table}[b]
\centering
\begin{tabular}{llcccccc}
Bands   & & $\chi^{(3)}$ & $Q_\mathrm{c}$ & & $\textbf{P}$ & Position & $N$\\ \hline
Group 1 & & $(0, 1)$     & 2/3  & & (2/3, 1/3) & $1b$ & 2\\
Group 2 & & $(0, 0)$     & 0     & & (0, 0) & $(1a)$  & 12\\
Group 3 & & $(-2, 2)$    & 0     & & (2/3, 1/3) & $1c$ & 4\\ 
\hline
All     & & $(-2, 3)$.   & 2/3   & & (1/3, 2/3) & & 18
\end{tabular}
\caption{$C_3$ Rotation indicator invariants of MoS$_2$ for the three groups of bands indicated in Fig. \ref{fig:bandstructure}, with their predicted corner charges $Q_\mathrm{c}$ (in units of $e=-|e|$), Wyckoff positions, polarization $\textbf{P}$ (units of $e/A$ in a basis of unit cell vectors) and the number of electrons $N$ occupying the Wyckoff position.}
\label{tab:MoS2invariants}
\end{table}

In the present work we have computed the symmetry indicator invariant \eqref{eq:symmetry_indicator} in MoS$_2$ and 70 other 2D insulators with the same space group taken from the computational 2D materials database (C2DB) \cite{Haastrup2018}. We have used the electronic structure software package GPAW \cite{Enkovaara2010, Larsen2017} using the PAW method and a plane wave basis. The computational parameters and lattice parameters are the same as those used in the C2DB. In Fig. \ref{fig:bandstructure}  we show the band structure of MoS$_2$ and in Tab. \ref{tab:MoS2invariants} we show the contributions to the symmetry indicator invariants from the different groups of bands. The total invariant is simply a sum of the different contributions and is given by $\chi^{(3)} = (-2, 3)$. The OAL phase thus emerges from the two upper valence bands that yields a Wannier center on the $1b$ Wyckoff position. The resulting Wannier functions originate from hybridization of $d$-orbitals from the three adjacent Molybdenum atoms \cite{Gibertini2015} and give rise to a fractional corner charge of $Q_\mathrm{c} = \frac{2e}{3}$ in MoS$_2$. It should be emphasized that unlike the case of quantum spin Hall insulators, the OAL phase is not driven by spin-orbit coupling and OAL phases are thus expected to be rather common in other transition metal dichalcogenides and dihalides with similar electronic structure. The four deep valence bands that are mainly composed of S $s$-orbitals have a non-vanishing symmetry indicator invariant and simply corresponds to four states located at the S atom (the $1c$ site). These states have trivial second order topology, but contributes to the topological polarization. With the choice of unit cell in Fig. \ref{fig:unitcell} the charges of the nuclei do not contribute to the polarization, which may thus be obtained by summing up the Wyckoff positions weighted by the occupancy stated in Tab. \ref{tab:MoS2invariants}.

\begin{figure}[tb]
\centering
\includegraphics[width=0.45\textwidth]{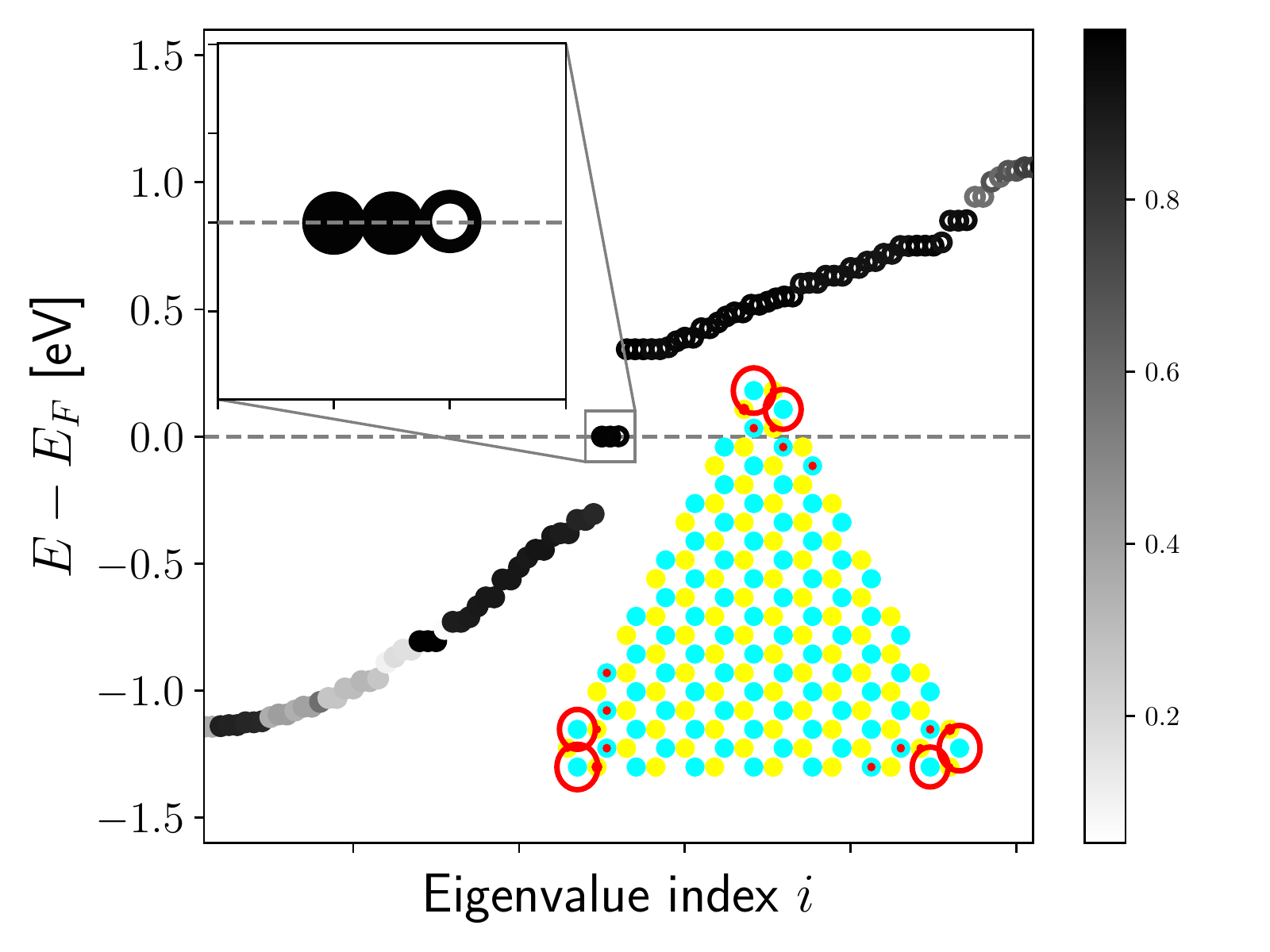}
\caption{Eigenvalue spectrum of a $C_3$ symmetric MoS$_2$ flake with armchair edges. The greyscale indicates the weight of a given state at edges (black is full edge localization). The top inset shows the three (Kramers degenerate) eigenvalues at the Fermi level, which are filled by four electrons at charge neutrality. At the bottom right we show the flake with the sum of the norm-squared wavefunctions of the states located at the Fermi level marked in red.}
\label{fig:armchair_flake}
\end{figure}

\begin{figure*}[tb]
    \centering
    \includegraphics[width=0.8\textwidth]{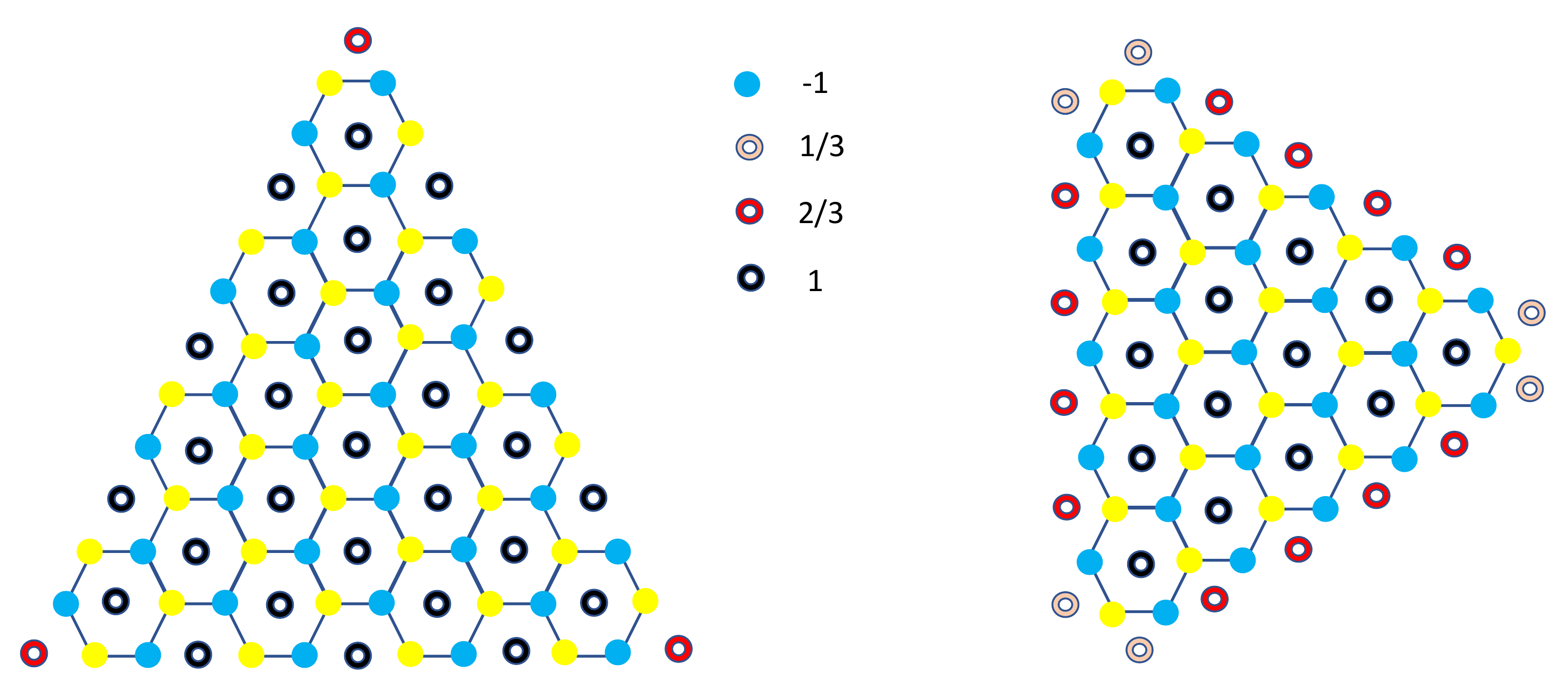}
    \caption{C$_3$ symmetric triangles with a topological polarization of (1/3, 2/3). The bulk structure has a single occupied state at 1b (center of hexagons) which is taken from the 1a position (blue atoms), while the yellow sites (1c) are neutral. Left: triangle with armchair edges, which are fully occupied, whereas the corners have fractional occupancy of 2/3. Right: triangle with zigzag edges where 1b sites at {\it both} corners and edges are fractionally occupied. The edges thus host metallic states that cannot be passivated.}
    \label{fig:MoS2_triangle_polar}
\end{figure*}

In order to verify the emergence of fractional corner charges we have constructed tight-binding models of triangular MoS$_2$ nanoflakes. The tight-binding parameters were obtained from the Kohn-Sham Hamiltonian of bulk MoS$_2$ in a Wannier function representation obtained with Wannier90 package \cite{Mostofi2008, Olsen2016a}. An example with armchair edges is shown in Fig. \ref{fig:armchair_flake} along with the eigenvalues colored according to weight on edge. We see that the spectrum is gapped except for three (Kramers degenerate) eigenvalues that are pinned to the Fermi level. The associated wavefunctions are localized at the corners of the triangle and becomes pinned because there are only four available electrons for filling the triangle to the neutrality point. The fractional corner charges thus arise as a consequence of this filling anomaly. We find that explicit integration of the charge density over each symmetry related sector yields exponentially localized corner charges of $Q_\mathrm{c}=2e/3$, when the degenerate levels are filled. At the neutrality point such flakes will exhibit anomalous polarizability \cite{Aihara2020} since a small electric field will break the symmetry and localize four states at two corners. This will give rise to a total flake dipole moment of $Q_\mathrm{c}L$, where $L$ is the height of the triangle. 

Triangular nanoparticles of MoS$_2$ have been synthesized on Au surfaces and characterized by STM \cite{bollinger1}. From the experiments it is clear that these nanoparticles exhibit zigzag edges terminated by S$_2$ dimers and theoretical predictions \cite{bollinger2} have verified that such nanoparticles are in general much more stable compared to armchair terminated nanoparticles. However, the zigzag edges always carry gapless states, which can be attributed to a non-trivial topological polarization orthogonal to the zigzag direction in MoS$_2$ \cite{Gibertini2015}. The metallic edge states thus cannot be passivated by adsorbates and the fractional corner charges will always be unobservable. In the supplementary material we show an example of a flake calculation with zigzag edges where the lack of a gap at the Fermi level is evident. We also present DFT calculations of band structures and potential profiles of MoS$_2$ nanoribbons, which explicitly shows that the bulk polarization gives rise to charge transfer between edges and that gapless bands at zigzag edges are enforced by the polarization.

It is perhaps not obvious that the bulk topological polarization have any physical consequences for nanotriangles, which are intrinsically non-polar due to symmetry. To illustrate that this is indeed the case we consider a simple model of a one-electron system in the OAL phase (occupation of the 1b site). Such a system is representative of MoS$_2$ and in Fig. \ref{fig:MoS2_triangle_polar} we show cartoons of nanotriangles where a single electron has been taken from the transition metal atom and placed in the 1b site. This corresponds to the case of $Q_\mathrm{c}=1/3$ (not taking Kramers degeneracy into account) and a polarization of $(1/3, 2/3)$. A symmetric construction in terms of Wannier charge centers is obtained by taken the electron originating from each Mo atom and distributing 1/3 at the three neighboring 1b (hollow) sites. In the case of the armchair edges this results in alternating hollow sites filled with 1/3 and 2/3 of an electron and these can be combined into integer occupations such that the edges become fully compensated and fractional charges only reside at corners. For the zigzag triangle, it is not possible to construct compensated edges in this way and the edge states become fractionally occupied and dispersive. The charge at a given edge is $2(N-1)/3$, where $N$ is the number of edge unit cells. Similarly, with a corner charge of $2/3$ any zigzag ribbon would acquire fractionally occupied edge states with an edge charge of $(N-1)/3$. Since a symmetry preserving addition of edge states can only change the site occupations by integer values there is no way to eliminate the dispersive edge states by adsorbates. For a one-electron model the fractional corner charges are associated with occupation of the 1b site, which will always lead to a topological polarization. The fact that the case of MoS$_2$ can be mapped to a one-band model where dispersive bands at zigzag edges are enforced is intimately connected to the non-trivial topological polarization. However, as will be shown below it is possible to find examples where such a mapping does not exist and this implies the existence of insulating states at {\it both} armchair and zigzag edges if the topological polarization vanishes.

\section{Computational screening of C2DB}
\begin{figure}[tb]
\centering
\includegraphics[width=0.45\textwidth]{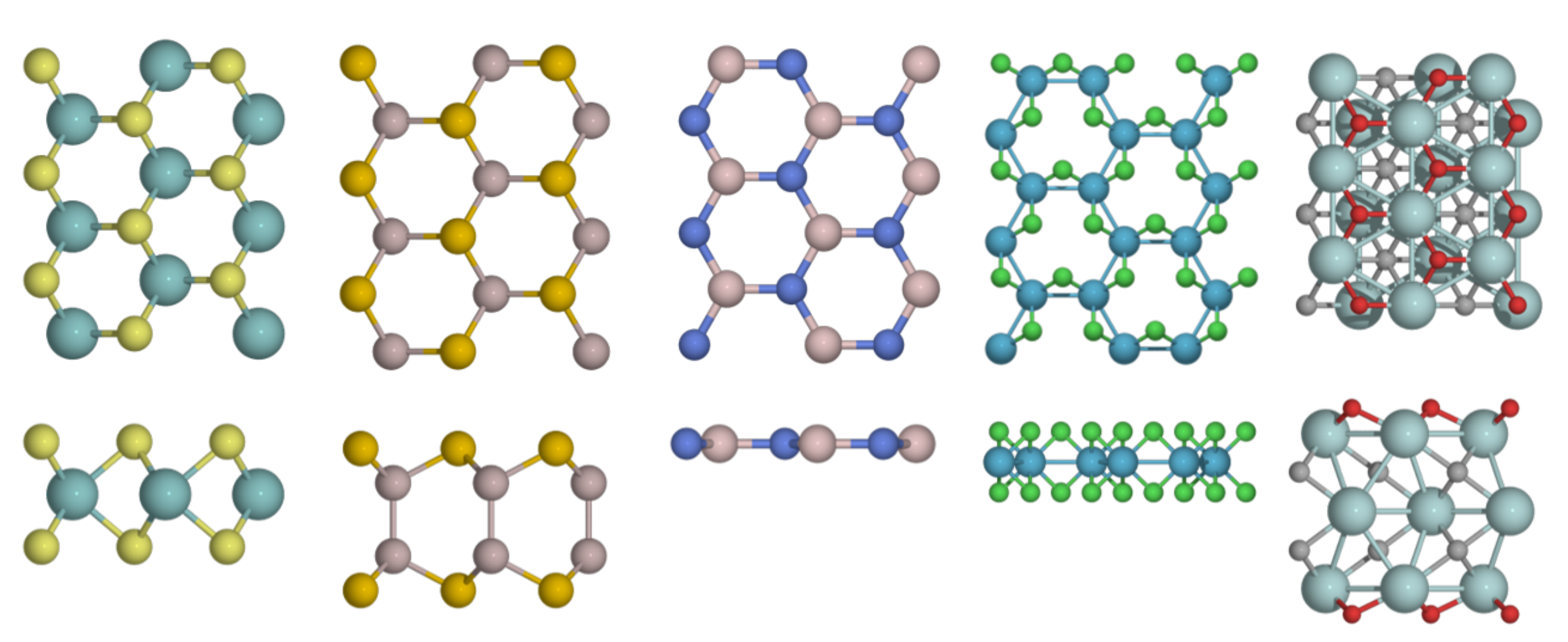}
\caption{Prototypical examples of 2D materials from the C2DB with space group $P\bar 6m2$. Left to right: MoS$_2$, Ga$_2$S$_2$, hBN, W$_2$Br$_6$ and Hf$_3$C$_2$O$_2$.}
\label{fig:structures}
\end{figure}
\begin{table}[tb]
\begin{tabular}{l|lllll}
             &  & $\chi^{(3)}$ & $Q_\mathrm{c}$ & $\textbf{P}$ & ID      \\ \hline
BN           &  & (0, 0)       & 0              & (0, 0)       & 186248  \\
Ga$_2$S$_2$  &  & (-2, 2)      & 0              & (2/3, 1/3)   & 635254  \\
Ga$_2$Se$_2$ &  & (-2, 2)      & 0              & (2/3, 1/3)   & 2002    \\
Ga$_2$Te$_2$ &  & (-2, 2)      & 0              & (2/3, 1/3)   & 43328   \\
GaN          &  & (0, 0)       & 0              & (0, 0)       & 159250  \\
In$_2$Se$_2$ &  & (-2, 2)      & 0              & (2/3, 1/3)   & 640503  \\
MoS$_2$      &  & (-2, 3)      & 2/3            & (1/3, 2/3)   & 38401   \\
MoSe$_2$     &  & (-2, 3)      & 2/3            & (1/3, 2/3)   & 49800   \\
MoTe$_2$     &  & (-2, 3)      & 2/3            & (1/3, 2/3)   & 15431   \\
WS$_2$       &  & (-2, 3)      & 2/3            & (1/3, 2/3)   & 202367  \\
WSe$_2$      &  & (-2, 3)      & 2/3            & (1/3, 2/3)   & 84182   \\
WTe$_2$      &  & (-2, 3)      & 2/3            & (1/3, 2/3)   & 653170  \\
ZrCl$_2$     &  & (-2, 3)      & 2/3            & (0, 0)       & 1530902
\end{tabular}
\caption{2D insulators that are experimentally known as van der Waals bonded 3D materials. The columns are the symmetry indicator $\chi^{(3)}$, fractional corner charges $Q_\mathrm{c}$ (in units of $e$), polarization $\textbf{P}$ and ICSD/COD indentifer (ID).}
\label{tab:maininvariants}
\end{table}

\begin{figure*}[tb]
\centering
\includegraphics[width=0.99\textwidth]{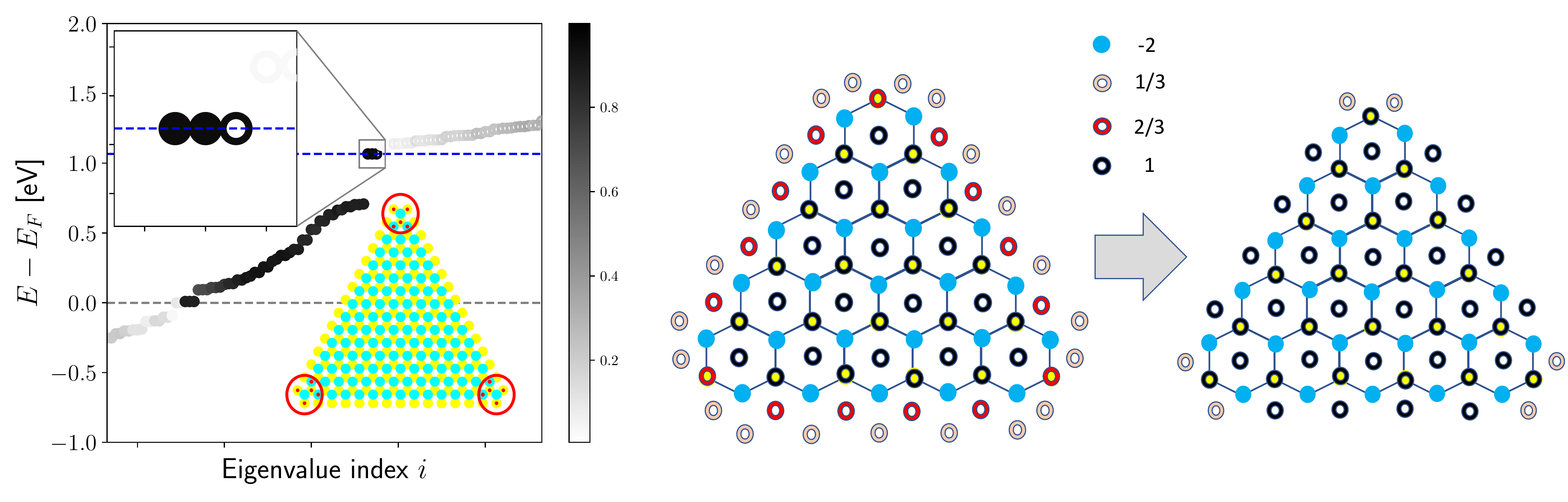}
\caption{Left: Eigenvalue spectrum of a $C_3$ symmetric TiCl$_2$ flake with Cl$_2$ terminated zigzag edges. The color indicates the edge localization of the states (dark is fully localized at edges). The Fermi level at zero energy (grey dashed line) corresponds to the bare flake and the Fermi level at 1.2 eV results from passivating all Cl dimers at edges with two electrons. The top inset show a zoom in on the three (Kramers degenerate) eigenvalues pinned to the Fermi level (in the edge-passivated system). Right: C$_3$ symmetric zigzag triangle with vanishing polarization. The bulk structure has one occupied state at 1b (center of hexagons) and one occupied state at 1c (Cl sites), which are both taken from the 1a position (Ti atoms). We first show the result of assigning 1/3 of an electron to all 1b and 1c sites adjacent to 1a and then show how then edge sites may be combined into integer occupied sites. This results in passivated edges and corner states with a fractional occupation of 2/3.}
\label{fig:zigzag_flake}
\end{figure*}
For 2D materials similar to MoS$_2$ the zigzag terminated nanotriangles are expected to be more stable than armchair terminated triangles as well and it is therefore pertinent to look for 2D materials that have non-trivial second order topology and associated fractional corner charges, but a trivial polarization. We have thus screened 71 stable materials from the C2DB \cite{Haastrup2018} with the space group $P\bar 6m2$ and time-reversal symmetry for non-trivial second order topology. In Fig. \ref{fig:structures} we show examples of the different prototypes included in the search and we have tabulated all results in the supplementary. To summarize, we find 28 materials in the OAL phase and 16 of these have vanishing polarization. In Table \ref{tab:maininvariants} we highlight invariants of experimentally known van der Waals materials within the criteria search space. Prototypes with similar valence occupation typically yield identical symmetry indicator invariants and polarization. We wish to emphasise the case of ZrCl$_2$, as we find it to have corner charges while having no polarization, thus being prime candidate for experimental verification in both C$_3$ symmetric triangles shown in Fig. \ref{fig:MoS2_triangle_polar}. MoX$_2$ and WX$_2$ (X = S, Se, Te) are isostructural to ZrCl$_2$, but due to the finite polarization, the corner charges are only accessible in the less stable armchair morphology. The cases of Ga$_2$X$_2$ (X = S, Se, Te) and In$_2$Se$_2$ have non-trivial polarization but no fractionalized corner states and thus only metallic edge states are expected in the zigzag structure. Lastly, BN and GaN are both trivial in term of polarization and corner charges. Bottom-up and Top-down approaches exists for experimental fabrication of triangular nanostructures. In the former approach, crystals are grown by conventional chemical vapor deposition methods, which generates atomically sharp structures in the most stable structure (zigzag edges). In contrast, anisotropic etching methods generate thin film structures with the least stable armchair edge. However, while the corner states are stable with respect to C$_3$ symmetric perturbations, there are still challenges to reduce roughness and thickness of samples \cite{danielsen2021super}.

For the compounds not presented in Table \ref{tab:maininvariants}, we find that materials with the same stoichiometry are rather similar in terms of the topology of individual bands. For example the band structure of HfTe$_2$ is highly similar to that of MoS$_2$ except there is two valence electrons less. The non-trivial bands are thus above the Fermi level, which yields a trivial topology and a polarization of $(2/3, 1/3)$ instead of $(1/3, 2/3)$. Another example is comprised by TiCl$_2$ which has the same number of valence electrons and fractional corner charges as MoS$_2$, but in that case the charge of the nuclei exactly cancels the electronic polarization, such that the total polarization vanishes (see supplementary for DFT calculations of TiCl$_2$ ribbons). TiCl$_2$ thus comprises a good example of a material where the corner charges might be observable in real (zigzag terminated) triangles by atomically resolved STM.
The material has not yet been synthesized experimentally, but is situated on the convex hull \cite{Haastrup2018} and is thus predicted to be thermodynamically stable (the heat of formation is -1.54 eV per unit cell). In Fig. \ref{fig:zigzag_flake} we show an example of the eigenvalue spectrum of a stable TiCl$_2$ triangle with zigzag edges. The band gap of bulk TiCl$_2$ is 0.9 eV and we again observe dispersive states in the bulk band gap that shift the corner localized states away from the Fermi level. In contrast to the case of MoS$_2$, however, the metallic edge states originate from dangling bonds associated with Cl dimers at the edge rather than bulk polarization and these may be passivated by appropriate adsorbates. For example, adding adsorbates that donate two electrons per Cl dimer shifts the Fermi level above the dispersive band in the gap and it becomes pinned to three (Kramers degenerate) eigenvalues at 1.2 eV (see Fig. \ref{fig:zigzag_flake}) that correspond to states localized at corners. This yields fractional corner charges of 2/3 (modulo two) and insulating edges, as is required for a unique STM signature in the zigzag morphology.  

Similar to the case of MoS$_2$ we can map the system onto a minimal model and consider how the occupation of different Wyckoff position may lead to fractional charges at edges and corners. However, the OAL phase combined with a vanishing polarization implies that one needs at least a two-band model to represent the system. We thus consider the case where two electrons are removed from the transition metal atom and one is added at the 1b site and another one at the 1c site. This yields a vanishing polarization and a fractional corner charge of $1/3$. We show this construction for a zigzag nanotriangle in Fig. \ref{fig:zigzag_flake}. For clarity, we first indicate a direct assignment of 1/3 of an electron at all 1b {\it and} 1b sites adjacent to Ti atoms (1a sites). This gives rise to alternating edge sites occupied by 1/3 and 2/3 of an electron, which may be combined to yield fully compensated edges and the final results is a zigzag triangle with fractional charges residing at corners only.

\section{Conclusion}
To conclude, we have calculated polarization and fractional corner charges of 71 stable 2D materials with three-fold rotational symmetry and found 16 materials that have an OAL phase {\it and} and a vanishing polarization. It has been argued that only these materials are relevant for experimental verification using STM\cite{helveg2000atomic}, since the fractional corner charges of OAL materials with non-trivial polarization will be obscured by edged states in nanotriangles with zigzag edges. We note that this was already concluded in the seminal work of Benalcazar et al \cite{Benalcazar2019}, but in that study the polarization and OAL was considered in the context of a single band and the non-trivial topological polarization is then always associated with an OAL phase. It should be emphasized that we have only included materials that are predicted to be dynamically and thermodynamically stable and all of the 16 materials could thus be relevant for experimental verification of second order topology. Of particular interest, however, are materials that are known to exist as bulk van der Waals structures, since these are expected to be directly exfoliable from bulk. In Tab. \ref{tab:maininvariants} we list all such materials (from the 71 materials) along with their identifier from the databases of experimental structures ICSD \cite{Allmann2007} or COD \cite{Graulis2011}. We find one material - ZrCl$_2$ - that has been experimentally characterized in bulk form \cite{doi:10.1021/ic50193a060} and is predicted to have an OAL phase as well trivial polarization. This material may thus comprise an optimal starting point for the observation of second order topology in 2D.

We thank Ivo Souza for fruitful discussions related to this work. T.O. and J.S. acknowledge support from the Villum foundation Grant No. 00029378.

\bibliography{bibliography}

\begin{thebibliography}{44}%
\makeatletter
\providecommand \@ifxundefined [1]{%
 \@ifx{#1\undefined}
}%
\providecommand \@ifnum [1]{%
 \ifnum #1\expandafter \@firstoftwo
 \else \expandafter \@secondoftwo
 \fi
}%
\providecommand \@ifx [1]{%
 \ifx #1\expandafter \@firstoftwo
 \else \expandafter \@secondoftwo
 \fi
}%
\providecommand \natexlab [1]{#1}%
\providecommand \enquote  [1]{``#1''}%
\providecommand \bibnamefont  [1]{#1}%
\providecommand \bibfnamefont [1]{#1}%
\providecommand \citenamefont [1]{#1}%
\providecommand \href@noop [0]{\@secondoftwo}%
\providecommand \href [0]{\begingroup \@sanitize@url \@href}%
\providecommand \@href[1]{\@@startlink{#1}\@@href}%
\providecommand \@@href[1]{\endgroup#1\@@endlink}%
\providecommand \@sanitize@url [0]{\catcode `\\12\catcode `\$12\catcode
  `\&12\catcode `\#12\catcode `\^12\catcode `\_12\catcode `\%12\relax}%
\providecommand \@@startlink[1]{}%
\providecommand \@@endlink[0]{}%
\providecommand \url  [0]{\begingroup\@sanitize@url \@url }%
\providecommand \@url [1]{\endgroup\@href {#1}{\urlprefix }}%
\providecommand \urlprefix  [0]{URL }%
\providecommand \Eprint [0]{\href }%
\providecommand \doibase [0]{https://doi.org/}%
\providecommand \selectlanguage [0]{\@gobble}%
\providecommand \bibinfo  [0]{\@secondoftwo}%
\providecommand \bibfield  [0]{\@secondoftwo}%
\providecommand \translation [1]{[#1]}%
\providecommand \BibitemOpen [0]{}%
\providecommand \bibitemStop [0]{}%
\providecommand \bibitemNoStop [0]{.\EOS\space}%
\providecommand \EOS [0]{\spacefactor3000\relax}%
\providecommand \BibitemShut  [1]{\csname bibitem#1\endcsname}%
\let\auto@bib@innerbib\@empty
\bibitem [{\citenamefont {Kane}\ and\ \citenamefont
  {Mele}(2005{\natexlab{a}})}]{Kane2005}%
  \BibitemOpen
  \bibfield  {author} {\bibinfo {author} {\bibfnamefont {C.~L.}\ \bibnamefont
  {Kane}}\ and\ \bibinfo {author} {\bibfnamefont {E.~J.}\ \bibnamefont
  {Mele}},\ }\href {https://doi.org/10.1103/PhysRevLett.95.146802} {\bibfield
  {journal} {\bibinfo  {journal} {Physical Review Letters}\ }\textbf {\bibinfo
  {volume} {95}},\ \bibinfo {pages} {146802} (\bibinfo {year}
  {2005}{\natexlab{a}})}\BibitemShut {NoStop}%
\bibitem [{\citenamefont {Kane}\ and\ \citenamefont
  {Mele}(2005{\natexlab{b}})}]{Kane2005a}%
  \BibitemOpen
  \bibfield  {author} {\bibinfo {author} {\bibfnamefont {C.~L.}\ \bibnamefont
  {Kane}}\ and\ \bibinfo {author} {\bibfnamefont {E.~J.}\ \bibnamefont
  {Mele}},\ }\href {https://doi.org/10.1103/PhysRevLett.95.226801} {\bibfield
  {journal} {\bibinfo  {journal} {Physical Review Letters}\ }\textbf {\bibinfo
  {volume} {95}},\ \bibinfo {pages} {226801} (\bibinfo {year}
  {2005}{\natexlab{b}})}\BibitemShut {NoStop}%
\bibitem [{\citenamefont {Bernevig}\ \emph {et~al.}(2006)\citenamefont
  {Bernevig}, \citenamefont {Hughes},\ and\ \citenamefont
  {Zhang}}]{Bernevig2006}%
  \BibitemOpen
  \bibfield  {author} {\bibinfo {author} {\bibfnamefont {B.~A.}\ \bibnamefont
  {Bernevig}}, \bibinfo {author} {\bibfnamefont {T.~L.}\ \bibnamefont
  {Hughes}},\ and\ \bibinfo {author} {\bibfnamefont {S.-C.}\ \bibnamefont
  {Zhang}},\ }\href {https://doi.org/10.1126/science.1133734} {\bibfield
  {journal} {\bibinfo  {journal} {Science}\ }\textbf {\bibinfo {volume}
  {314}},\ \bibinfo {pages} {1757} (\bibinfo {year} {2006})}\BibitemShut
  {NoStop}%
\bibitem [{\citenamefont {Konig}\ \emph {et~al.}(2007)\citenamefont {Konig},
  \citenamefont {Wiedmann}, \citenamefont {Brune}, \citenamefont {Roth},
  \citenamefont {Buhmann}, \citenamefont {Molenkamp}, \citenamefont {Qi},\ and\
  \citenamefont {Zhang}}]{Konig2007}%
  \BibitemOpen
  \bibfield  {author} {\bibinfo {author} {\bibfnamefont {M.}~\bibnamefont
  {Konig}}, \bibinfo {author} {\bibfnamefont {S.}~\bibnamefont {Wiedmann}},
  \bibinfo {author} {\bibfnamefont {C.}~\bibnamefont {Brune}}, \bibinfo
  {author} {\bibfnamefont {A.}~\bibnamefont {Roth}}, \bibinfo {author}
  {\bibfnamefont {H.}~\bibnamefont {Buhmann}}, \bibinfo {author} {\bibfnamefont
  {L.~W.}\ \bibnamefont {Molenkamp}}, \bibinfo {author} {\bibfnamefont {X.-L.}\
  \bibnamefont {Qi}},\ and\ \bibinfo {author} {\bibfnamefont {S.-C.}\
  \bibnamefont {Zhang}},\ }\href {https://doi.org/10.1126/science.1148047}
  {\bibfield  {journal} {\bibinfo  {journal} {Science}\ }\textbf {\bibinfo
  {volume} {318}},\ \bibinfo {pages} {766} (\bibinfo {year}
  {2007})}\BibitemShut {NoStop}%
\bibitem [{\citenamefont {Hasan}\ and\ \citenamefont {Kane}(2010)}]{Hasan2010}%
  \BibitemOpen
  \bibfield  {author} {\bibinfo {author} {\bibfnamefont {M.~Z.}\ \bibnamefont
  {Hasan}}\ and\ \bibinfo {author} {\bibfnamefont {C.~L.}\ \bibnamefont
  {Kane}},\ }\href {https://doi.org/10.1103/RevModPhys.82.3045} {\bibfield
  {journal} {\bibinfo  {journal} {Reviews of Modern Physics}\ }\textbf
  {\bibinfo {volume} {82}},\ \bibinfo {pages} {3045} (\bibinfo {year}
  {2010})}\BibitemShut {NoStop}%
\bibitem [{\citenamefont {Moore}(2010)}]{Moore2010}%
  \BibitemOpen
  \bibfield  {author} {\bibinfo {author} {\bibfnamefont {J.~E.}\ \bibnamefont
  {Moore}},\ }\href {https://doi.org/10.1038/nature08916} {\bibfield  {journal}
  {\bibinfo  {journal} {Nature}\ }\textbf {\bibinfo {volume} {464}},\ \bibinfo
  {pages} {194} (\bibinfo {year} {2010})}\BibitemShut {NoStop}%
\bibitem [{\citenamefont {Br{\"{u}}ne}\ \emph {et~al.}(2012)\citenamefont
  {Br{\"{u}}ne}, \citenamefont {Roth}, \citenamefont {Buhmann}, \citenamefont
  {Hankiewicz}, \citenamefont {Molenkamp}, \citenamefont {Maciejko},
  \citenamefont {Qi},\ and\ \citenamefont {Zhang}}]{Brune2012}%
  \BibitemOpen
  \bibfield  {author} {\bibinfo {author} {\bibfnamefont {C.}~\bibnamefont
  {Br{\"{u}}ne}}, \bibinfo {author} {\bibfnamefont {A.}~\bibnamefont {Roth}},
  \bibinfo {author} {\bibfnamefont {H.}~\bibnamefont {Buhmann}}, \bibinfo
  {author} {\bibfnamefont {E.~M.}\ \bibnamefont {Hankiewicz}}, \bibinfo
  {author} {\bibfnamefont {L.~W.}\ \bibnamefont {Molenkamp}}, \bibinfo {author}
  {\bibfnamefont {J.}~\bibnamefont {Maciejko}}, \bibinfo {author}
  {\bibfnamefont {X.-L.}\ \bibnamefont {Qi}},\ and\ \bibinfo {author}
  {\bibfnamefont {S.-C.}\ \bibnamefont {Zhang}},\ }\href
  {https://doi.org/10.1038/nphys2322} {\bibfield  {journal} {\bibinfo
  {journal} {Nature Physics}\ }\textbf {\bibinfo {volume} {8}},\ \bibinfo
  {pages} {485} (\bibinfo {year} {2012})},\ \Eprint
  {https://arxiv.org/abs/1107.0585} {1107.0585} \BibitemShut {NoStop}%
\bibitem [{\citenamefont {Qian}\ \emph {et~al.}(2014)\citenamefont {Qian},
  \citenamefont {Liu}, \citenamefont {Fu},\ and\ \citenamefont
  {Li}}]{Qian2014}%
  \BibitemOpen
  \bibfield  {author} {\bibinfo {author} {\bibfnamefont {X.}~\bibnamefont
  {Qian}}, \bibinfo {author} {\bibfnamefont {J.}~\bibnamefont {Liu}}, \bibinfo
  {author} {\bibfnamefont {L.}~\bibnamefont {Fu}},\ and\ \bibinfo {author}
  {\bibfnamefont {J.}~\bibnamefont {Li}},\ }\href
  {https://doi.org/10.1126/science.1256815} {\bibfield  {journal} {\bibinfo
  {journal} {Science}\ }\textbf {\bibinfo {volume} {346}},\ \bibinfo {pages}
  {1344} (\bibinfo {year} {2014})}\BibitemShut {NoStop}%
\bibitem [{\citenamefont {Olsen}\ \emph {et~al.}(2019)\citenamefont {Olsen},
  \citenamefont {Andersen}, \citenamefont {Okugawa}, \citenamefont {Torelli},
  \citenamefont {Deilmann},\ and\ \citenamefont {Thygesen}}]{Olsen2018}%
  \BibitemOpen
  \bibfield  {author} {\bibinfo {author} {\bibfnamefont {T.}~\bibnamefont
  {Olsen}}, \bibinfo {author} {\bibfnamefont {E.}~\bibnamefont {Andersen}},
  \bibinfo {author} {\bibfnamefont {T.}~\bibnamefont {Okugawa}}, \bibinfo
  {author} {\bibfnamefont {D.}~\bibnamefont {Torelli}}, \bibinfo {author}
  {\bibfnamefont {T.}~\bibnamefont {Deilmann}},\ and\ \bibinfo {author}
  {\bibfnamefont {K.~S.}\ \bibnamefont {Thygesen}},\ }\href
  {https://doi.org/10.1103/PhysRevMaterials.3.024005} {\bibfield  {journal}
  {\bibinfo  {journal} {Physical Review Materials}\ }\textbf {\bibinfo {volume}
  {3}},\ \bibinfo {pages} {024005} (\bibinfo {year} {2019})},\ \Eprint
  {https://arxiv.org/abs/1812.06666} {arXiv:1812.06666} \BibitemShut {NoStop}%
\bibitem [{\citenamefont {Fu}\ and\ \citenamefont {Kane}(2007)}]{Fu2007}%
  \BibitemOpen
  \bibfield  {author} {\bibinfo {author} {\bibfnamefont {L.}~\bibnamefont
  {Fu}}\ and\ \bibinfo {author} {\bibfnamefont {C.~L.}\ \bibnamefont {Kane}},\
  }\href {https://doi.org/10.1103/PhysRevB.76.045302} {\bibfield  {journal}
  {\bibinfo  {journal} {Physical Review B}\ }\textbf {\bibinfo {volume} {76}},\
  \bibinfo {pages} {045302} (\bibinfo {year} {2007})}\BibitemShut {NoStop}%
\bibitem [{\citenamefont {Fu}\ \emph {et~al.}(2007)\citenamefont {Fu},
  \citenamefont {Kane},\ and\ \citenamefont {Mele}}]{Fu2007a}%
  \BibitemOpen
  \bibfield  {author} {\bibinfo {author} {\bibfnamefont {L.}~\bibnamefont
  {Fu}}, \bibinfo {author} {\bibfnamefont {C.~L.}\ \bibnamefont {Kane}},\ and\
  \bibinfo {author} {\bibfnamefont {E.~J.}\ \bibnamefont {Mele}},\ }\href
  {https://doi.org/10.1103/PhysRevLett.98.106803} {\bibfield  {journal}
  {\bibinfo  {journal} {Physical Review Letters}\ }\textbf {\bibinfo {volume}
  {98}},\ \bibinfo {pages} {106803} (\bibinfo {year} {2007})}\BibitemShut
  {NoStop}%
\bibitem [{\citenamefont {Qi}\ \emph {et~al.}(2008)\citenamefont {Qi},
  \citenamefont {Hughes},\ and\ \citenamefont {Zhang}}]{Qi2008}%
  \BibitemOpen
  \bibfield  {author} {\bibinfo {author} {\bibfnamefont {X.-L.}\ \bibnamefont
  {Qi}}, \bibinfo {author} {\bibfnamefont {T.~L.}\ \bibnamefont {Hughes}},\
  and\ \bibinfo {author} {\bibfnamefont {S.-C.}\ \bibnamefont {Zhang}},\ }\href
  {https://doi.org/10.1103/PhysRevB.78.195424} {\bibfield  {journal} {\bibinfo
  {journal} {Physical Review B}\ }\textbf {\bibinfo {volume} {78}},\ \bibinfo
  {pages} {195424} (\bibinfo {year} {2008})}\BibitemShut {NoStop}%
\bibitem [{\citenamefont {Zhang}\ \emph {et~al.}(2009)\citenamefont {Zhang},
  \citenamefont {Liu}, \citenamefont {Qi}, \citenamefont {Dai}, \citenamefont
  {Fang},\ and\ \citenamefont {Zhang}}]{Zhang2009}%
  \BibitemOpen
  \bibfield  {author} {\bibinfo {author} {\bibfnamefont {H.}~\bibnamefont
  {Zhang}}, \bibinfo {author} {\bibfnamefont {C.-X.}\ \bibnamefont {Liu}},
  \bibinfo {author} {\bibfnamefont {X.-L.}\ \bibnamefont {Qi}}, \bibinfo
  {author} {\bibfnamefont {X.}~\bibnamefont {Dai}}, \bibinfo {author}
  {\bibfnamefont {Z.}~\bibnamefont {Fang}},\ and\ \bibinfo {author}
  {\bibfnamefont {S.-C.}\ \bibnamefont {Zhang}},\ }\href
  {https://doi.org/10.1038/nphys1270} {\bibfield  {journal} {\bibinfo
  {journal} {Nature Physics}\ }\textbf {\bibinfo {volume} {5}},\ \bibinfo
  {pages} {438} (\bibinfo {year} {2009})}\BibitemShut {NoStop}%
\bibitem [{\citenamefont {Fu}(2011)}]{Fu2011}%
  \BibitemOpen
  \bibfield  {author} {\bibinfo {author} {\bibfnamefont {L.}~\bibnamefont
  {Fu}},\ }\href {https://doi.org/10.1103/PhysRevLett.106.106802} {\bibfield
  {journal} {\bibinfo  {journal} {Physical Review Letters}\ }\textbf {\bibinfo
  {volume} {106}},\ \bibinfo {pages} {106802} (\bibinfo {year}
  {2011})}\BibitemShut {NoStop}%
\bibitem [{\citenamefont {Hsieh}\ \emph {et~al.}(2012)\citenamefont {Hsieh},
  \citenamefont {Lin}, \citenamefont {Liu}, \citenamefont {Duan}, \citenamefont
  {Bansil},\ and\ \citenamefont {Fu}}]{Hsieh2012}%
  \BibitemOpen
  \bibfield  {author} {\bibinfo {author} {\bibfnamefont {T.~H.}\ \bibnamefont
  {Hsieh}}, \bibinfo {author} {\bibfnamefont {H.}~\bibnamefont {Lin}}, \bibinfo
  {author} {\bibfnamefont {J.}~\bibnamefont {Liu}}, \bibinfo {author}
  {\bibfnamefont {W.}~\bibnamefont {Duan}}, \bibinfo {author} {\bibfnamefont
  {A.}~\bibnamefont {Bansil}},\ and\ \bibinfo {author} {\bibfnamefont
  {L.}~\bibnamefont {Fu}},\ }\href {https://doi.org/10.1038/ncomms1969}
  {\bibfield  {journal} {\bibinfo  {journal} {Nature Communications}\ }\textbf
  {\bibinfo {volume} {3}},\ \bibinfo {pages} {982} (\bibinfo {year} {2012})},\
  \Eprint {https://arxiv.org/abs/1202.1003} {1202.1003} \BibitemShut {NoStop}%
\bibitem [{\citenamefont {Kargarian}\ and\ \citenamefont
  {Fiete}(2013)}]{Kargarian2013}%
  \BibitemOpen
  \bibfield  {author} {\bibinfo {author} {\bibfnamefont {M.}~\bibnamefont
  {Kargarian}}\ and\ \bibinfo {author} {\bibfnamefont {G.~A.}\ \bibnamefont
  {Fiete}},\ }\href {https://doi.org/10.1103/PhysRevLett.110.156403} {\bibfield
   {journal} {\bibinfo  {journal} {Physical Review Letters}\ }\textbf {\bibinfo
  {volume} {110}},\ \bibinfo {pages} {1} (\bibinfo {year} {2013})}\BibitemShut
  {NoStop}%
\bibitem [{\citenamefont {Liu}\ \emph {et~al.}(2014)\citenamefont {Liu},
  \citenamefont {Hsieh}, \citenamefont {Wei}, \citenamefont {Duan},
  \citenamefont {Moodera},\ and\ \citenamefont {Fu}}]{Liu2014}%
  \BibitemOpen
  \bibfield  {author} {\bibinfo {author} {\bibfnamefont {J.}~\bibnamefont
  {Liu}}, \bibinfo {author} {\bibfnamefont {T.~H.}\ \bibnamefont {Hsieh}},
  \bibinfo {author} {\bibfnamefont {P.}~\bibnamefont {Wei}}, \bibinfo {author}
  {\bibfnamefont {W.}~\bibnamefont {Duan}}, \bibinfo {author} {\bibfnamefont
  {J.}~\bibnamefont {Moodera}},\ and\ \bibinfo {author} {\bibfnamefont
  {L.}~\bibnamefont {Fu}},\ }\href {https://doi.org/10.1038/nmat3828}
  {\bibfield  {journal} {\bibinfo  {journal} {Nature Materials}\ }\textbf
  {\bibinfo {volume} {13}},\ \bibinfo {pages} {178} (\bibinfo {year}
  {2014})}\BibitemShut {NoStop}%
\bibitem [{\citenamefont {Fei}\ \emph {et~al.}(2017)\citenamefont {Fei},
  \citenamefont {Palomaki}, \citenamefont {Wu}, \citenamefont {Zhao},
  \citenamefont {Cai}, \citenamefont {Sun}, \citenamefont {Nguyen},
  \citenamefont {Finney}, \citenamefont {Xu},\ and\ \citenamefont
  {Cobden}}]{Fei2017}%
  \BibitemOpen
  \bibfield  {author} {\bibinfo {author} {\bibfnamefont {Z.}~\bibnamefont
  {Fei}}, \bibinfo {author} {\bibfnamefont {T.}~\bibnamefont {Palomaki}},
  \bibinfo {author} {\bibfnamefont {S.}~\bibnamefont {Wu}}, \bibinfo {author}
  {\bibfnamefont {W.}~\bibnamefont {Zhao}}, \bibinfo {author} {\bibfnamefont
  {X.}~\bibnamefont {Cai}}, \bibinfo {author} {\bibfnamefont {B.}~\bibnamefont
  {Sun}}, \bibinfo {author} {\bibfnamefont {P.}~\bibnamefont {Nguyen}},
  \bibinfo {author} {\bibfnamefont {J.}~\bibnamefont {Finney}}, \bibinfo
  {author} {\bibfnamefont {X.}~\bibnamefont {Xu}},\ and\ \bibinfo {author}
  {\bibfnamefont {D.~H.}\ \bibnamefont {Cobden}},\ }\href
  {https://doi.org/10.1038/nphys4091} {\bibfield  {journal} {\bibinfo
  {journal} {Nature Physics}\ }\textbf {\bibinfo {volume} {13}},\ \bibinfo
  {pages} {677} (\bibinfo {year} {2017})},\ \Eprint
  {https://arxiv.org/abs/1610.07924} {1610.07924} \BibitemShut {NoStop}%
\bibitem [{\citenamefont {Ugeda}\ \emph {et~al.}(2018)\citenamefont {Ugeda},
  \citenamefont {Pulkin}, \citenamefont {Tang}, \citenamefont {Ryu},
  \citenamefont {Wu}, \citenamefont {Zhang}, \citenamefont {Wong},
  \citenamefont {Pedramrazi}, \citenamefont {Mart{\'{i}}n-Recio}, \citenamefont
  {Chen}, \citenamefont {Wang}, \citenamefont {Shen}, \citenamefont {Mo},
  \citenamefont {Yazyev},\ and\ \citenamefont {Crommie}}]{Ugeda2018}%
  \BibitemOpen
  \bibfield  {author} {\bibinfo {author} {\bibfnamefont {M.~M.}\ \bibnamefont
  {Ugeda}}, \bibinfo {author} {\bibfnamefont {A.}~\bibnamefont {Pulkin}},
  \bibinfo {author} {\bibfnamefont {S.}~\bibnamefont {Tang}}, \bibinfo {author}
  {\bibfnamefont {H.}~\bibnamefont {Ryu}}, \bibinfo {author} {\bibfnamefont
  {Q.}~\bibnamefont {Wu}}, \bibinfo {author} {\bibfnamefont {Y.}~\bibnamefont
  {Zhang}}, \bibinfo {author} {\bibfnamefont {D.}~\bibnamefont {Wong}},
  \bibinfo {author} {\bibfnamefont {Z.}~\bibnamefont {Pedramrazi}}, \bibinfo
  {author} {\bibfnamefont {A.}~\bibnamefont {Mart{\'{i}}n-Recio}}, \bibinfo
  {author} {\bibfnamefont {Y.}~\bibnamefont {Chen}}, \bibinfo {author}
  {\bibfnamefont {F.}~\bibnamefont {Wang}}, \bibinfo {author} {\bibfnamefont
  {Z.-X.}\ \bibnamefont {Shen}}, \bibinfo {author} {\bibfnamefont {S.-K.}\
  \bibnamefont {Mo}}, \bibinfo {author} {\bibfnamefont {O.~V.}\ \bibnamefont
  {Yazyev}},\ and\ \bibinfo {author} {\bibfnamefont {M.~F.}\ \bibnamefont
  {Crommie}},\ }\href {https://doi.org/10.1038/s41467-018-05672-w} {\bibfield
  {journal} {\bibinfo  {journal} {Nature Communications}\ }\textbf {\bibinfo
  {volume} {9}},\ \bibinfo {pages} {3401} (\bibinfo {year} {2018})}\BibitemShut
  {NoStop}%
\bibitem [{\citenamefont {Benalcazar}\ \emph
  {et~al.}(2017{\natexlab{a}})\citenamefont {Benalcazar}, \citenamefont
  {Bernevig},\ and\ \citenamefont {Hughes}}]{Benalcazar2017a}%
  \BibitemOpen
  \bibfield  {author} {\bibinfo {author} {\bibfnamefont {W.~A.}\ \bibnamefont
  {Benalcazar}}, \bibinfo {author} {\bibfnamefont {B.~A.}\ \bibnamefont
  {Bernevig}},\ and\ \bibinfo {author} {\bibfnamefont {T.~L.}\ \bibnamefont
  {Hughes}},\ }\href {https://doi.org/10.1103/PhysRevB.96.245115} {\bibfield
  {journal} {\bibinfo  {journal} {Physical Review B}\ }\textbf {\bibinfo
  {volume} {96}},\ \bibinfo {pages} {245115} (\bibinfo {year}
  {2017}{\natexlab{a}})}\BibitemShut {NoStop}%
\bibitem [{\citenamefont {Su}\ \emph {et~al.}(1979)\citenamefont {Su},
  \citenamefont {Schrieffer},\ and\ \citenamefont {Heeger}}]{Su1979}%
  \BibitemOpen
  \bibfield  {author} {\bibinfo {author} {\bibfnamefont {W.~P.}\ \bibnamefont
  {Su}}, \bibinfo {author} {\bibfnamefont {J.~R.}\ \bibnamefont {Schrieffer}},\
  and\ \bibinfo {author} {\bibfnamefont {A.~J.}\ \bibnamefont {Heeger}},\
  }\href {https://doi.org/10.1103/PhysRevLett.42.1698} {\bibfield  {journal}
  {\bibinfo  {journal} {Physical Review Letters}\ }\textbf {\bibinfo {volume}
  {42}},\ \bibinfo {pages} {1698} (\bibinfo {year} {1979})}\BibitemShut
  {NoStop}%
\bibitem [{\citenamefont {Gibertini}\ and\ \citenamefont
  {Marzari}(2015)}]{Gibertini2015}%
  \BibitemOpen
  \bibfield  {author} {\bibinfo {author} {\bibfnamefont {M.}~\bibnamefont
  {Gibertini}}\ and\ \bibinfo {author} {\bibfnamefont {N.}~\bibnamefont
  {Marzari}},\ }\href {https://doi.org/10.1021/acs.nanolett.5b02834} {\bibfield
   {journal} {\bibinfo  {journal} {Nano Letters}\ }\textbf {\bibinfo {volume}
  {15}},\ \bibinfo {pages} {6229} (\bibinfo {year} {2015})}\BibitemShut
  {NoStop}%
\bibitem [{\citenamefont {Benalcazar}\ \emph
  {et~al.}(2017{\natexlab{b}})\citenamefont {Benalcazar}, \citenamefont
  {Bernevig},\ and\ \citenamefont {Hughes}}]{Benalcazar2017b}%
  \BibitemOpen
  \bibfield  {author} {\bibinfo {author} {\bibfnamefont {W.~A.}\ \bibnamefont
  {Benalcazar}}, \bibinfo {author} {\bibfnamefont {B.~A.}\ \bibnamefont
  {Bernevig}},\ and\ \bibinfo {author} {\bibfnamefont {T.~L.}\ \bibnamefont
  {Hughes}},\ }\href {https://doi.org/10.1126/science.aah6442} {\bibfield
  {journal} {\bibinfo  {journal} {Science}\ }\textbf {\bibinfo {volume}
  {357}},\ \bibinfo {pages} {61} (\bibinfo {year}
  {2017}{\natexlab{b}})}\BibitemShut {NoStop}%
\bibitem [{\citenamefont {Benalcazar}\ \emph {et~al.}(2019)\citenamefont
  {Benalcazar}, \citenamefont {Li},\ and\ \citenamefont
  {Hughes}}]{Benalcazar2019}%
  \BibitemOpen
  \bibfield  {author} {\bibinfo {author} {\bibfnamefont {W.~A.}\ \bibnamefont
  {Benalcazar}}, \bibinfo {author} {\bibfnamefont {T.}~\bibnamefont {Li}},\
  and\ \bibinfo {author} {\bibfnamefont {T.~L.}\ \bibnamefont {Hughes}},\
  }\href {https://doi.org/10.1103/PhysRevB.99.245151} {\bibfield  {journal}
  {\bibinfo  {journal} {Physical Review B}\ }\textbf {\bibinfo {volume} {99}},\
  \bibinfo {pages} {245151} (\bibinfo {year} {2019})}\BibitemShut {NoStop}%
\bibitem [{\citenamefont {Song}\ \emph {et~al.}(2017)\citenamefont {Song},
  \citenamefont {Fang},\ and\ \citenamefont {Fang}}]{Song2017b}%
  \BibitemOpen
  \bibfield  {author} {\bibinfo {author} {\bibfnamefont {Z.}~\bibnamefont
  {Song}}, \bibinfo {author} {\bibfnamefont {Z.}~\bibnamefont {Fang}},\ and\
  \bibinfo {author} {\bibfnamefont {C.}~\bibnamefont {Fang}},\ }\href
  {https://doi.org/10.1103/PhysRevLett.119.246402} {\bibfield  {journal}
  {\bibinfo  {journal} {Physical Review Letters}\ }\textbf {\bibinfo {volume}
  {119}},\ \bibinfo {pages} {246402} (\bibinfo {year} {2017})}\BibitemShut
  {NoStop}%
\bibitem [{\citenamefont {Langbehn}\ \emph {et~al.}(2017)\citenamefont
  {Langbehn}, \citenamefont {Peng}, \citenamefont {Trifunovic}, \citenamefont
  {von Oppen},\ and\ \citenamefont {Brouwer}}]{Langbehn2017b}%
  \BibitemOpen
  \bibfield  {author} {\bibinfo {author} {\bibfnamefont {J.}~\bibnamefont
  {Langbehn}}, \bibinfo {author} {\bibfnamefont {Y.}~\bibnamefont {Peng}},
  \bibinfo {author} {\bibfnamefont {L.}~\bibnamefont {Trifunovic}}, \bibinfo
  {author} {\bibfnamefont {F.}~\bibnamefont {von Oppen}},\ and\ \bibinfo
  {author} {\bibfnamefont {P.~W.}\ \bibnamefont {Brouwer}},\ }\href
  {https://doi.org/10.1103/PhysRevLett.119.246401} {\bibfield  {journal}
  {\bibinfo  {journal} {Physical Review Letters}\ }\textbf {\bibinfo {volume}
  {119}},\ \bibinfo {pages} {246401} (\bibinfo {year} {2017})}\BibitemShut
  {NoStop}%
\bibitem [{\citenamefont {Schindler}\ \emph
  {et~al.}(2018{\natexlab{a}})\citenamefont {Schindler}, \citenamefont {Cook},
  \citenamefont {Vergniory}, \citenamefont {Wang}, \citenamefont {Parkin},
  \citenamefont {Bernevig},\ and\ \citenamefont {Neupert}}]{Schindler2018a}%
  \BibitemOpen
  \bibfield  {author} {\bibinfo {author} {\bibfnamefont {F.}~\bibnamefont
  {Schindler}}, \bibinfo {author} {\bibfnamefont {A.~M.}\ \bibnamefont {Cook}},
  \bibinfo {author} {\bibfnamefont {M.~G.}\ \bibnamefont {Vergniory}}, \bibinfo
  {author} {\bibfnamefont {Z.}~\bibnamefont {Wang}}, \bibinfo {author}
  {\bibfnamefont {S.~S.~P.}\ \bibnamefont {Parkin}}, \bibinfo {author}
  {\bibfnamefont {B.~A.}\ \bibnamefont {Bernevig}},\ and\ \bibinfo {author}
  {\bibfnamefont {T.}~\bibnamefont {Neupert}},\ }\href
  {https://doi.org/10.1126/sciadv.aat0346} {\bibfield  {journal} {\bibinfo
  {journal} {Science Advances}\ }\textbf {\bibinfo {volume} {4}},\ \bibinfo
  {pages} {0346} (\bibinfo {year} {2018}{\natexlab{a}})}\BibitemShut {NoStop}%
\bibitem [{\citenamefont {Schindler}\ \emph
  {et~al.}(2018{\natexlab{b}})\citenamefont {Schindler}, \citenamefont {Wang},
  \citenamefont {Vergniory}, \citenamefont {Cook}, \citenamefont {Murani},
  \citenamefont {Sengupta}, \citenamefont {Kasumov}, \citenamefont {Deblock},
  \citenamefont {Jeon}, \citenamefont {Drozdov}, \citenamefont {Bouchiat},
  \citenamefont {Gu{\'{e}}ron}, \citenamefont {Yazdani}, \citenamefont
  {Bernevig},\ and\ \citenamefont {Neupert}}]{Schindler2018b}%
  \BibitemOpen
  \bibfield  {author} {\bibinfo {author} {\bibfnamefont {F.}~\bibnamefont
  {Schindler}}, \bibinfo {author} {\bibfnamefont {Z.}~\bibnamefont {Wang}},
  \bibinfo {author} {\bibfnamefont {M.~G.}\ \bibnamefont {Vergniory}}, \bibinfo
  {author} {\bibfnamefont {A.~M.}\ \bibnamefont {Cook}}, \bibinfo {author}
  {\bibfnamefont {A.}~\bibnamefont {Murani}}, \bibinfo {author} {\bibfnamefont
  {S.}~\bibnamefont {Sengupta}}, \bibinfo {author} {\bibfnamefont {A.~Y.}\
  \bibnamefont {Kasumov}}, \bibinfo {author} {\bibfnamefont {R.}~\bibnamefont
  {Deblock}}, \bibinfo {author} {\bibfnamefont {S.}~\bibnamefont {Jeon}},
  \bibinfo {author} {\bibfnamefont {I.}~\bibnamefont {Drozdov}}, \bibinfo
  {author} {\bibfnamefont {H.}~\bibnamefont {Bouchiat}}, \bibinfo {author}
  {\bibfnamefont {S.}~\bibnamefont {Gu{\'{e}}ron}}, \bibinfo {author}
  {\bibfnamefont {A.}~\bibnamefont {Yazdani}}, \bibinfo {author} {\bibfnamefont
  {B.~A.}\ \bibnamefont {Bernevig}},\ and\ \bibinfo {author} {\bibfnamefont
  {T.}~\bibnamefont {Neupert}},\ }\href
  {https://doi.org/10.1038/s41567-018-0224-7} {\bibfield  {journal} {\bibinfo
  {journal} {Nature Physics}\ }\textbf {\bibinfo {volume} {14}},\ \bibinfo
  {pages} {918} (\bibinfo {year} {2018}{\natexlab{b}})}\BibitemShut {NoStop}%
\bibitem [{\citenamefont {Qian}\ \emph {et~al.}(2022)\citenamefont {Qian},
  \citenamefont {Liu}, \citenamefont {Liu},\ and\ \citenamefont
  {Yao}}]{Qian2021}%
  \BibitemOpen
  \bibfield  {author} {\bibinfo {author} {\bibfnamefont {S.}~\bibnamefont
  {Qian}}, \bibinfo {author} {\bibfnamefont {G.-B.}\ \bibnamefont {Liu}},
  \bibinfo {author} {\bibfnamefont {C.-C.}\ \bibnamefont {Liu}},\ and\ \bibinfo
  {author} {\bibfnamefont {Y.}~\bibnamefont {Yao}},\ }\href
  {https://doi.org/10.1103/PhysRevB.105.045417} {\bibfield  {journal} {\bibinfo
   {journal} {Phys. Rev. B}\ }\textbf {\bibinfo {volume} {105}},\ \bibinfo
  {pages} {045417} (\bibinfo {year} {2022})}\BibitemShut {NoStop}%
\bibitem [{\citenamefont {Zeng}\ \emph {et~al.}(2021)\citenamefont {Zeng},
  \citenamefont {Liu}, \citenamefont {Jiang}, \citenamefont {Sun},\ and\
  \citenamefont {Xie}}]{Zeng2021}%
  \BibitemOpen
  \bibfield  {author} {\bibinfo {author} {\bibfnamefont {J.}~\bibnamefont
  {Zeng}}, \bibinfo {author} {\bibfnamefont {H.}~\bibnamefont {Liu}}, \bibinfo
  {author} {\bibfnamefont {H.}~\bibnamefont {Jiang}}, \bibinfo {author}
  {\bibfnamefont {Q.-F.}\ \bibnamefont {Sun}},\ and\ \bibinfo {author}
  {\bibfnamefont {X.~C.}\ \bibnamefont {Xie}},\ }\href
  {https://doi.org/10.1103/PhysRevB.104.L161108} {\bibfield  {journal}
  {\bibinfo  {journal} {Phys. Rev. B}\ }\textbf {\bibinfo {volume} {104}},\
  \bibinfo {pages} {L161108} (\bibinfo {year} {2021})}\BibitemShut {NoStop}%
\bibitem [{\citenamefont {Helveg}\ \emph {et~al.}(2000)\citenamefont {Helveg},
  \citenamefont {Lauritsen}, \citenamefont {L{\ae}gsgaard}, \citenamefont
  {Stensgaard}, \citenamefont {N{\o}rskov}, \citenamefont {Clausen},
  \citenamefont {Tops{\o}e},\ and\ \citenamefont
  {Besenbacher}}]{helveg2000atomic}%
  \BibitemOpen
  \bibfield  {author} {\bibinfo {author} {\bibfnamefont {S.}~\bibnamefont
  {Helveg}}, \bibinfo {author} {\bibfnamefont {J.~V.}\ \bibnamefont
  {Lauritsen}}, \bibinfo {author} {\bibfnamefont {E.}~\bibnamefont
  {L{\ae}gsgaard}}, \bibinfo {author} {\bibfnamefont {I.}~\bibnamefont
  {Stensgaard}}, \bibinfo {author} {\bibfnamefont {J.~K.}\ \bibnamefont
  {N{\o}rskov}}, \bibinfo {author} {\bibfnamefont {B.}~\bibnamefont {Clausen}},
  \bibinfo {author} {\bibfnamefont {H.}~\bibnamefont {Tops{\o}e}},\ and\
  \bibinfo {author} {\bibfnamefont {F.}~\bibnamefont {Besenbacher}},\
  }\href@noop {} {\bibfield  {journal} {\bibinfo  {journal} {Physical review
  letters}\ }\textbf {\bibinfo {volume} {84}},\ \bibinfo {pages} {951}
  (\bibinfo {year} {2000})}\BibitemShut {NoStop}%
\bibitem [{\citenamefont {Schindler}\ \emph {et~al.}(2019)\citenamefont
  {Schindler}, \citenamefont {Brzezi{\'{n}}ska}, \citenamefont {Benalcazar},
  \citenamefont {Iraola}, \citenamefont {Bouhon}, \citenamefont {Tsirkin},
  \citenamefont {Vergniory},\ and\ \citenamefont {Neupert}}]{Schindler2019}%
  \BibitemOpen
  \bibfield  {author} {\bibinfo {author} {\bibfnamefont {F.}~\bibnamefont
  {Schindler}}, \bibinfo {author} {\bibfnamefont {M.}~\bibnamefont
  {Brzezi{\'{n}}ska}}, \bibinfo {author} {\bibfnamefont {W.~A.}\ \bibnamefont
  {Benalcazar}}, \bibinfo {author} {\bibfnamefont {M.}~\bibnamefont {Iraola}},
  \bibinfo {author} {\bibfnamefont {A.}~\bibnamefont {Bouhon}}, \bibinfo
  {author} {\bibfnamefont {S.~S.}\ \bibnamefont {Tsirkin}}, \bibinfo {author}
  {\bibfnamefont {M.~G.}\ \bibnamefont {Vergniory}},\ and\ \bibinfo {author}
  {\bibfnamefont {T.}~\bibnamefont {Neupert}},\ }\href
  {https://doi.org/10.1103/PhysRevResearch.1.033074} {\bibfield  {journal}
  {\bibinfo  {journal} {Physical Review Research}\ }\textbf {\bibinfo {volume}
  {1}},\ \bibinfo {pages} {033074} (\bibinfo {year} {2019})}\BibitemShut
  {NoStop}%
\bibitem [{\citenamefont {Haastrup}\ \emph {et~al.}(2018)\citenamefont
  {Haastrup}, \citenamefont {Strange}, \citenamefont {Pandey}, \citenamefont
  {Deilmann}, \citenamefont {Schmidt}, \citenamefont {Hinsche}, \citenamefont
  {Gjerding}, \citenamefont {Torelli}, \citenamefont {Larsen}, \citenamefont
  {Riis-Jensen}, \citenamefont {Gath}, \citenamefont {Jacobsen}, \citenamefont
  {Mortensen}, \citenamefont {Olsen},\ and\ \citenamefont
  {Thygesen}}]{Haastrup2018}%
  \BibitemOpen
  \bibfield  {author} {\bibinfo {author} {\bibfnamefont {S.}~\bibnamefont
  {Haastrup}}, \bibinfo {author} {\bibfnamefont {M.}~\bibnamefont {Strange}},
  \bibinfo {author} {\bibfnamefont {M.}~\bibnamefont {Pandey}}, \bibinfo
  {author} {\bibfnamefont {T.}~\bibnamefont {Deilmann}}, \bibinfo {author}
  {\bibfnamefont {P.~S.}\ \bibnamefont {Schmidt}}, \bibinfo {author}
  {\bibfnamefont {N.~F.}\ \bibnamefont {Hinsche}}, \bibinfo {author}
  {\bibfnamefont {M.~N.}\ \bibnamefont {Gjerding}}, \bibinfo {author}
  {\bibfnamefont {D.}~\bibnamefont {Torelli}}, \bibinfo {author} {\bibfnamefont
  {P.~M.}\ \bibnamefont {Larsen}}, \bibinfo {author} {\bibfnamefont {A.~C.}\
  \bibnamefont {Riis-Jensen}}, \bibinfo {author} {\bibfnamefont
  {J.}~\bibnamefont {Gath}}, \bibinfo {author} {\bibfnamefont {K.~W.}\
  \bibnamefont {Jacobsen}}, \bibinfo {author} {\bibfnamefont {J.~J.}\
  \bibnamefont {Mortensen}}, \bibinfo {author} {\bibfnamefont {T.}~\bibnamefont
  {Olsen}},\ and\ \bibinfo {author} {\bibfnamefont {K.~S.}\ \bibnamefont
  {Thygesen}},\ }\href {https://doi.org/10.1088/2053-1583/aacfc1} {\bibfield
  {journal} {\bibinfo  {journal} {2D Materials}\ }\textbf {\bibinfo {volume}
  {5}},\ \bibinfo {pages} {042002} (\bibinfo {year} {2018})},\ \Eprint
  {https://arxiv.org/abs/1806.03173} {arXiv:1806.03173} \BibitemShut {NoStop}%
\bibitem [{\citenamefont {Enkovaara}\ \emph {et~al.}(2010)\citenamefont
  {Enkovaara}, \citenamefont {Rostgaard}, \citenamefont {Mortensen},
  \citenamefont {Chen}, \citenamefont {Du{\l}ak}, \citenamefont {Ferrighi},
  \citenamefont {Gavnholt}, \citenamefont {Glinsvad}, \citenamefont {Haikola},
  \citenamefont {Hansen}, \citenamefont {Kristoffersen}, \citenamefont
  {Kuisma}, \citenamefont {Larsen}, \citenamefont {Lehtovaara}, \citenamefont
  {Ljungberg}, \citenamefont {Lopez-Acevedo}, \citenamefont {Moses},
  \citenamefont {Ojanen}, \citenamefont {Olsen}, \citenamefont {Petzold},
  \citenamefont {Romero}, \citenamefont {Stausholm-M{\o}ller}, \citenamefont
  {Strange}, \citenamefont {Tritsaris}, \citenamefont {Vanin}, \citenamefont
  {Walter}, \citenamefont {Hammer}, \citenamefont {H{\"{a}}kkinen},
  \citenamefont {Madsen}, \citenamefont {Nieminen}, \citenamefont {N{\o}rskov},
  \citenamefont {Puska}, \citenamefont {Rantala}, \citenamefont {Schi{\o}tz},
  \citenamefont {Thygesen}, \citenamefont {Jacobsen},\ and\ \citenamefont
  {Others}}]{Enkovaara2010}%
  \BibitemOpen
  \bibfield  {author} {\bibinfo {author} {\bibfnamefont {J.}~\bibnamefont
  {Enkovaara}}, \bibinfo {author} {\bibfnamefont {C.}~\bibnamefont
  {Rostgaard}}, \bibinfo {author} {\bibfnamefont {J.~J.}\ \bibnamefont
  {Mortensen}}, \bibinfo {author} {\bibfnamefont {J.}~\bibnamefont {Chen}},
  \bibinfo {author} {\bibfnamefont {M.}~\bibnamefont {Du{\l}ak}}, \bibinfo
  {author} {\bibfnamefont {L.}~\bibnamefont {Ferrighi}}, \bibinfo {author}
  {\bibfnamefont {J.}~\bibnamefont {Gavnholt}}, \bibinfo {author}
  {\bibfnamefont {C.}~\bibnamefont {Glinsvad}}, \bibinfo {author}
  {\bibfnamefont {V.}~\bibnamefont {Haikola}}, \bibinfo {author} {\bibfnamefont
  {H.~a.}\ \bibnamefont {Hansen}}, \bibinfo {author} {\bibfnamefont {H.~H.}\
  \bibnamefont {Kristoffersen}}, \bibinfo {author} {\bibfnamefont
  {M.}~\bibnamefont {Kuisma}}, \bibinfo {author} {\bibfnamefont {a.~H.}\
  \bibnamefont {Larsen}}, \bibinfo {author} {\bibfnamefont {L.}~\bibnamefont
  {Lehtovaara}}, \bibinfo {author} {\bibfnamefont {M.}~\bibnamefont
  {Ljungberg}}, \bibinfo {author} {\bibfnamefont {O.}~\bibnamefont
  {Lopez-Acevedo}}, \bibinfo {author} {\bibfnamefont {P.~G.}\ \bibnamefont
  {Moses}}, \bibinfo {author} {\bibfnamefont {J.}~\bibnamefont {Ojanen}},
  \bibinfo {author} {\bibfnamefont {T.}~\bibnamefont {Olsen}}, \bibinfo
  {author} {\bibfnamefont {V.}~\bibnamefont {Petzold}}, \bibinfo {author}
  {\bibfnamefont {N.~a.}\ \bibnamefont {Romero}}, \bibinfo {author}
  {\bibfnamefont {J.}~\bibnamefont {Stausholm-M{\o}ller}}, \bibinfo {author}
  {\bibfnamefont {M.}~\bibnamefont {Strange}}, \bibinfo {author} {\bibfnamefont
  {G.~a.}\ \bibnamefont {Tritsaris}}, \bibinfo {author} {\bibfnamefont
  {M.}~\bibnamefont {Vanin}}, \bibinfo {author} {\bibfnamefont
  {M.}~\bibnamefont {Walter}}, \bibinfo {author} {\bibfnamefont
  {B.}~\bibnamefont {Hammer}}, \bibinfo {author} {\bibfnamefont
  {H.}~\bibnamefont {H{\"{a}}kkinen}}, \bibinfo {author} {\bibfnamefont
  {G.~K.~H.}\ \bibnamefont {Madsen}}, \bibinfo {author} {\bibfnamefont {R.~M.}\
  \bibnamefont {Nieminen}}, \bibinfo {author} {\bibfnamefont {J.~K.}\
  \bibnamefont {N{\o}rskov}}, \bibinfo {author} {\bibfnamefont
  {M.}~\bibnamefont {Puska}}, \bibinfo {author} {\bibfnamefont {T.~T.}\
  \bibnamefont {Rantala}}, \bibinfo {author} {\bibfnamefont {J.}~\bibnamefont
  {Schi{\o}tz}}, \bibinfo {author} {\bibfnamefont {K.~S.}\ \bibnamefont
  {Thygesen}}, \bibinfo {author} {\bibfnamefont {K.~W.}\ \bibnamefont
  {Jacobsen}},\ and\ \bibinfo {author} {\bibnamefont {Others}},\ }\href
  {https://doi.org/10.1088/0953-8984/22/25/253202} {\bibfield  {journal}
  {\bibinfo  {journal} {Journal of Physics: Condensed Matter}\ }\textbf
  {\bibinfo {volume} {22}},\ \bibinfo {pages} {253202} (\bibinfo {year}
  {2010})}\BibitemShut {NoStop}%
\bibitem [{\citenamefont {{Hjorth Larsen}}\ \emph {et~al.}(2017)\citenamefont
  {{Hjorth Larsen}}, \citenamefont {{J{\o}rgen Mortensen}}, \citenamefont
  {Blomqvist}, \citenamefont {Castelli}, \citenamefont {Christensen},
  \citenamefont {Du{\l}ak}, \citenamefont {Friis}, \citenamefont {Groves},
  \citenamefont {Hammer}, \citenamefont {Hargus}, \citenamefont {Hermes},
  \citenamefont {Jennings}, \citenamefont {{Bjerre Jensen}}, \citenamefont
  {Kermode}, \citenamefont {Kitchin}, \citenamefont {{Leonhard Kolsbjerg}},
  \citenamefont {Kubal}, \citenamefont {Kaasbjerg}, \citenamefont {Lysgaard},
  \citenamefont {{Bergmann Maronsson}}, \citenamefont {Maxson}, \citenamefont
  {Olsen}, \citenamefont {Pastewka}, \citenamefont {Peterson}, \citenamefont
  {Rostgaard}, \citenamefont {Schi{\o}tz}, \citenamefont {Sch{\"{u}}tt},
  \citenamefont {Strange}, \citenamefont {Thygesen}, \citenamefont {Vegge},
  \citenamefont {Vilhelmsen}, \citenamefont {Walter}, \citenamefont {Zeng},\
  and\ \citenamefont {Jacobsen}}]{Larsen2017}%
  \BibitemOpen
  \bibfield  {author} {\bibinfo {author} {\bibfnamefont {A.}~\bibnamefont
  {{Hjorth Larsen}}}, \bibinfo {author} {\bibfnamefont {J.}~\bibnamefont
  {{J{\o}rgen Mortensen}}}, \bibinfo {author} {\bibfnamefont {J.}~\bibnamefont
  {Blomqvist}}, \bibinfo {author} {\bibfnamefont {I.~E.}\ \bibnamefont
  {Castelli}}, \bibinfo {author} {\bibfnamefont {R.}~\bibnamefont
  {Christensen}}, \bibinfo {author} {\bibfnamefont {M.}~\bibnamefont
  {Du{\l}ak}}, \bibinfo {author} {\bibfnamefont {J.}~\bibnamefont {Friis}},
  \bibinfo {author} {\bibfnamefont {M.~N.}\ \bibnamefont {Groves}}, \bibinfo
  {author} {\bibfnamefont {B.}~\bibnamefont {Hammer}}, \bibinfo {author}
  {\bibfnamefont {C.}~\bibnamefont {Hargus}}, \bibinfo {author} {\bibfnamefont
  {E.~D.}\ \bibnamefont {Hermes}}, \bibinfo {author} {\bibfnamefont {P.~C.}\
  \bibnamefont {Jennings}}, \bibinfo {author} {\bibfnamefont {P.}~\bibnamefont
  {{Bjerre Jensen}}}, \bibinfo {author} {\bibfnamefont {J.}~\bibnamefont
  {Kermode}}, \bibinfo {author} {\bibfnamefont {J.~R.}\ \bibnamefont
  {Kitchin}}, \bibinfo {author} {\bibfnamefont {E.}~\bibnamefont {{Leonhard
  Kolsbjerg}}}, \bibinfo {author} {\bibfnamefont {J.}~\bibnamefont {Kubal}},
  \bibinfo {author} {\bibfnamefont {K.}~\bibnamefont {Kaasbjerg}}, \bibinfo
  {author} {\bibfnamefont {S.}~\bibnamefont {Lysgaard}}, \bibinfo {author}
  {\bibfnamefont {J.}~\bibnamefont {{Bergmann Maronsson}}}, \bibinfo {author}
  {\bibfnamefont {T.}~\bibnamefont {Maxson}}, \bibinfo {author} {\bibfnamefont
  {T.}~\bibnamefont {Olsen}}, \bibinfo {author} {\bibfnamefont
  {L.}~\bibnamefont {Pastewka}}, \bibinfo {author} {\bibfnamefont
  {A.}~\bibnamefont {Peterson}}, \bibinfo {author} {\bibfnamefont
  {C.}~\bibnamefont {Rostgaard}}, \bibinfo {author} {\bibfnamefont
  {J.}~\bibnamefont {Schi{\o}tz}}, \bibinfo {author} {\bibfnamefont
  {O.}~\bibnamefont {Sch{\"{u}}tt}}, \bibinfo {author} {\bibfnamefont
  {M.}~\bibnamefont {Strange}}, \bibinfo {author} {\bibfnamefont {K.~S.}\
  \bibnamefont {Thygesen}}, \bibinfo {author} {\bibfnamefont {T.}~\bibnamefont
  {Vegge}}, \bibinfo {author} {\bibfnamefont {L.}~\bibnamefont {Vilhelmsen}},
  \bibinfo {author} {\bibfnamefont {M.}~\bibnamefont {Walter}}, \bibinfo
  {author} {\bibfnamefont {Z.}~\bibnamefont {Zeng}},\ and\ \bibinfo {author}
  {\bibfnamefont {K.~W.}\ \bibnamefont {Jacobsen}},\ }\href
  {https://doi.org/10.1088/1361-648X/aa680e} {\bibfield  {journal} {\bibinfo
  {journal} {Journal of Physics: Condensed Matter}\ }\textbf {\bibinfo {volume}
  {29}},\ \bibinfo {pages} {273002} (\bibinfo {year} {2017})}\BibitemShut
  {NoStop}%
\bibitem [{\citenamefont {Mostofi}\ \emph {et~al.}(2008)\citenamefont
  {Mostofi}, \citenamefont {Yates}, \citenamefont {Lee}, \citenamefont {Souza},
  \citenamefont {Vanderbilt},\ and\ \citenamefont {Marzari}}]{Mostofi2008}%
  \BibitemOpen
  \bibfield  {author} {\bibinfo {author} {\bibfnamefont {A.~A.}\ \bibnamefont
  {Mostofi}}, \bibinfo {author} {\bibfnamefont {J.~R.}\ \bibnamefont {Yates}},
  \bibinfo {author} {\bibfnamefont {Y.-S.}\ \bibnamefont {Lee}}, \bibinfo
  {author} {\bibfnamefont {I.}~\bibnamefont {Souza}}, \bibinfo {author}
  {\bibfnamefont {D.}~\bibnamefont {Vanderbilt}},\ and\ \bibinfo {author}
  {\bibfnamefont {N.}~\bibnamefont {Marzari}},\ }\href
  {https://doi.org/10.1016/j.cpc.2007.11.016} {\bibfield  {journal} {\bibinfo
  {journal} {Computer Physics Communications}\ }\textbf {\bibinfo {volume}
  {178}},\ \bibinfo {pages} {685} (\bibinfo {year} {2008})}\BibitemShut
  {NoStop}%
\bibitem [{\citenamefont {Olsen}(2016)}]{Olsen2016a}%
  \BibitemOpen
  \bibfield  {author} {\bibinfo {author} {\bibfnamefont {T.}~\bibnamefont
  {Olsen}},\ }\href {https://doi.org/10.1103/PhysRevB.94.235106} {\bibfield
  {journal} {\bibinfo  {journal} {Physical Review B}\ }\textbf {\bibinfo
  {volume} {94}},\ \bibinfo {pages} {235106} (\bibinfo {year}
  {2016})}\BibitemShut {NoStop}%
\bibitem [{\citenamefont {Aihara}\ \emph {et~al.}(2020)\citenamefont {Aihara},
  \citenamefont {Hirayama},\ and\ \citenamefont {Murakami}}]{Aihara2020}%
  \BibitemOpen
  \bibfield  {author} {\bibinfo {author} {\bibfnamefont {Y.}~\bibnamefont
  {Aihara}}, \bibinfo {author} {\bibfnamefont {M.}~\bibnamefont {Hirayama}},\
  and\ \bibinfo {author} {\bibfnamefont {S.}~\bibnamefont {Murakami}},\ }\href
  {https://doi.org/10.1103/PhysRevResearch.2.033224} {\bibfield  {journal}
  {\bibinfo  {journal} {Physical Review Research}\ }\textbf {\bibinfo {volume}
  {2}},\ \bibinfo {pages} {033224} (\bibinfo {year} {2020})},\ \Eprint
  {https://arxiv.org/abs/2006.03605} {arXiv:2006.03605} \BibitemShut {NoStop}%
\bibitem [{\citenamefont {Bollinger}\ \emph {et~al.}(2001)\citenamefont
  {Bollinger}, \citenamefont {Lauritsen}, \citenamefont {Jacobsen},
  \citenamefont {N{\o}rskov}, \citenamefont {Helveg},\ and\ \citenamefont
  {Besenbacher}}]{bollinger1}%
  \BibitemOpen
  \bibfield  {author} {\bibinfo {author} {\bibfnamefont {M.~V.}\ \bibnamefont
  {Bollinger}}, \bibinfo {author} {\bibfnamefont {J.~V.}\ \bibnamefont
  {Lauritsen}}, \bibinfo {author} {\bibfnamefont {K.~W.}\ \bibnamefont
  {Jacobsen}}, \bibinfo {author} {\bibfnamefont {J.~K.}\ \bibnamefont
  {N{\o}rskov}}, \bibinfo {author} {\bibfnamefont {S.}~\bibnamefont {Helveg}},\
  and\ \bibinfo {author} {\bibfnamefont {F.}~\bibnamefont {Besenbacher}},\
  }\href {https://doi.org/10.1103/PhysRevLett.87.196803} {\bibfield  {journal}
  {\bibinfo  {journal} {Physical Review Letters}\ }\textbf {\bibinfo {volume}
  {87}},\ \bibinfo {pages} {196803} (\bibinfo {year} {2001})}\BibitemShut
  {NoStop}%
\bibitem [{\citenamefont {Bollinger}\ \emph {et~al.}(2003)\citenamefont
  {Bollinger}, \citenamefont {Jacobsen},\ and\ \citenamefont
  {N{\o}rskov}}]{bollinger2}%
  \BibitemOpen
  \bibfield  {author} {\bibinfo {author} {\bibfnamefont {M.~V.}\ \bibnamefont
  {Bollinger}}, \bibinfo {author} {\bibfnamefont {K.~W.}\ \bibnamefont
  {Jacobsen}},\ and\ \bibinfo {author} {\bibfnamefont {J.~K.}\ \bibnamefont
  {N{\o}rskov}},\ }\href {https://doi.org/10.1103/PhysRevB.67.085410}
  {\bibfield  {journal} {\bibinfo  {journal} {Physical Review B}\ }\textbf
  {\bibinfo {volume} {67}},\ \bibinfo {pages} {085410} (\bibinfo {year}
  {2003})}\BibitemShut {NoStop}%
\bibitem [{\citenamefont {Danielsen}\ \emph {et~al.}(2021)\citenamefont
  {Danielsen}, \citenamefont {Lyksborg-Andersen}, \citenamefont {Nielsen},
  \citenamefont {Jessen}, \citenamefont {Booth}, \citenamefont {Doan},
  \citenamefont {Zhou}, \citenamefont {B{\o}ggild},\ and\ \citenamefont
  {Gammelgaard}}]{danielsen2021super}%
  \BibitemOpen
  \bibfield  {author} {\bibinfo {author} {\bibfnamefont {D.~R.}\ \bibnamefont
  {Danielsen}}, \bibinfo {author} {\bibfnamefont {A.}~\bibnamefont
  {Lyksborg-Andersen}}, \bibinfo {author} {\bibfnamefont {K.~E.}\ \bibnamefont
  {Nielsen}}, \bibinfo {author} {\bibfnamefont {B.~S.}\ \bibnamefont {Jessen}},
  \bibinfo {author} {\bibfnamefont {T.~J.}\ \bibnamefont {Booth}}, \bibinfo
  {author} {\bibfnamefont {M.-H.}\ \bibnamefont {Doan}}, \bibinfo {author}
  {\bibfnamefont {Y.}~\bibnamefont {Zhou}}, \bibinfo {author} {\bibfnamefont
  {P.}~\bibnamefont {B{\o}ggild}},\ and\ \bibinfo {author} {\bibfnamefont
  {L.}~\bibnamefont {Gammelgaard}},\ }\href@noop {} {\bibfield  {journal}
  {\bibinfo  {journal} {ACS Applied Materials \& Interfaces}\ }\textbf
  {\bibinfo {volume} {13}},\ \bibinfo {pages} {41886} (\bibinfo {year}
  {2021})}\BibitemShut {NoStop}%
\bibitem [{\citenamefont {Allmann}\ and\ \citenamefont
  {Hinek}(2007)}]{Allmann2007}%
  \BibitemOpen
  \bibfield  {author} {\bibinfo {author} {\bibfnamefont {R.}~\bibnamefont
  {Allmann}}\ and\ \bibinfo {author} {\bibfnamefont {R.}~\bibnamefont
  {Hinek}},\ }\href {https://doi.org/10.1107/S0108767307038081} {\bibfield
  {journal} {\bibinfo  {journal} {Acta Crystallogr. Sect. A Found.
  Crystallogr.}\ }\textbf {\bibinfo {volume} {63}},\ \bibinfo {pages} {412}
  (\bibinfo {year} {2007})}\BibitemShut {NoStop}%
\bibitem [{\citenamefont {Gra{\v{z}}ulis}\ \emph {et~al.}(2011)\citenamefont
  {Gra{\v{z}}ulis}, \citenamefont {Da{\v{s}}kevi{\v{c}}}, \citenamefont
  {Merkys}, \citenamefont {Chateigner}, \citenamefont {Lutterotti},
  \citenamefont {Quir{\'{o}}s}, \citenamefont {Serebryanaya}, \citenamefont
  {Moeck}, \citenamefont {Downs},\ and\ \citenamefont {Bail}}]{Graulis2011}%
  \BibitemOpen
  \bibfield  {author} {\bibinfo {author} {\bibfnamefont {S.}~\bibnamefont
  {Gra{\v{z}}ulis}}, \bibinfo {author} {\bibfnamefont {A.}~\bibnamefont
  {Da{\v{s}}kevi{\v{c}}}}, \bibinfo {author} {\bibfnamefont {A.}~\bibnamefont
  {Merkys}}, \bibinfo {author} {\bibfnamefont {D.}~\bibnamefont {Chateigner}},
  \bibinfo {author} {\bibfnamefont {L.}~\bibnamefont {Lutterotti}}, \bibinfo
  {author} {\bibfnamefont {M.}~\bibnamefont {Quir{\'{o}}s}}, \bibinfo {author}
  {\bibfnamefont {N.~R.}\ \bibnamefont {Serebryanaya}}, \bibinfo {author}
  {\bibfnamefont {P.}~\bibnamefont {Moeck}}, \bibinfo {author} {\bibfnamefont
  {R.~T.}\ \bibnamefont {Downs}},\ and\ \bibinfo {author} {\bibfnamefont
  {A.~L.}\ \bibnamefont {Bail}},\ }\href {https://doi.org/10.1093/nar/gkr900}
  {\bibfield  {journal} {\bibinfo  {journal} {Nucleic Acids Research}\ }\textbf
  {\bibinfo {volume} {40}},\ \bibinfo {pages} {D420} (\bibinfo {year}
  {2011})}\BibitemShut {NoStop}%
\bibitem [{\citenamefont {Cisar}\ \emph {et~al.}(1979)\citenamefont {Cisar},
  \citenamefont {Corbett},\ and\ \citenamefont
  {Daake}}]{doi:10.1021/ic50193a060}%
  \BibitemOpen
  \bibfield  {author} {\bibinfo {author} {\bibfnamefont {A.}~\bibnamefont
  {Cisar}}, \bibinfo {author} {\bibfnamefont {J.~D.}\ \bibnamefont {Corbett}},\
  and\ \bibinfo {author} {\bibfnamefont {R.~L.}\ \bibnamefont {Daake}},\ }\href
  {https://doi.org/10.1021/ic50193a060} {\bibfield  {journal} {\bibinfo
  {journal} {Inorganic Chemistry}\ }\textbf {\bibinfo {volume} {18}},\ \bibinfo
  {pages} {836} (\bibinfo {year} {1979})}\BibitemShut {NoStop}%
\end{thebibliography}%

\end{document}